\documentclass[preprint,aps,showpacs,showkeys,nofootinbib]{revtex4}
\usepackage{amsmath}
\usepackage{graphicx}
\usepackage{hyperref}
\usepackage{natbib} \begin{document}
\title{Lepton flavor violating decays $Z\rightarrow l^{\pm}_{i}l^{\mp}_{j}$ in the B-L Supersymmetric Standard Model}
\author{Jia-Peng Huo$^{1,2,3}$\footnote{qq2868371800@126.com},
Xing-Xing Dong$^{1,2,3}$\footnote{dongxx@hbu.edu.cn},
Jiao Ma$^{1,2,3}$\footnote{majiaojiao65@126.com},
Shu-Min Zhao$^{1,2,3}$\footnote{zhaosm@hbu.edu.cn},
Cai Guo$^{1,2,3}$\footnote{guocai0322@163.com},
Hai-Bin Zhang$^{1,2,3}$\footnote{hbzhang@hbu.edu.cn},
Jin-Lei Yang$^{1,2,3}$\footnote{jlyang@hbu.edu.cn},
Tai-Fu Feng$^{1,2,3,4}$\footnote{fengtf@hbu.edu.cn}}

\affiliation{$^1$ College of Physics Science and Technology, Hebei University, Baoding, 071002, China\\
$^2$ Hebei Key Laboratory of High-precision Computation and Application of Quantum Field Theory, Baoding, 071002, China\\
$^3$ Hebei Research Center of the Basic Discipline for Computational Physics, Baoding, 071002, China\\
$^4$ Department of Physics, Chongqing University, Chongqing, 401331, China}

\begin{abstract}
  Lepton flavor violation (LFV) represents a clear new physics (NP) signal beyond the
standard model (SM). In this paper, we study LFV decays $Z\rightarrow l^{\pm}_{i}l^{\mp}_{j}$ in the B-L Supersymmetric Standard Model(B-LSSM). We calculate these processes separately in the mass eigenstate basis and the electroweak interaction basis, and the latter adopt the mass insertion approximation (MIA) method. The MIA clearly shows the effect of parameters on the LFV decays $Z\rightarrow l^{\pm}_{i}l^{\mp}_{j}$ in the analytic level, which provides a new way for us to analyze the LFV processes. At the same time, the corresponding constraints from the LFV decays $l^{-}_{j} \rightarrow l^{-}_{i} \gamma$ and  $(g-2)_{\mu}$  are considered to analyze the numerical results.
\end{abstract}


 \maketitle

 \section{Introduction}
In the SM, there are two different Yukawa matrices $Y_u$ and $Y_d$ in the quark sector, both involving the left-handed doublet $Q_{L}$ of the quark field, which makes them impossible to diagonalize in the same basis at the same time\cite{Calibbi:2017uvl}. This means that flavor violation exists in the quark sector and is controlled by the CKM matrix\cite{Cabibbo:1963yz,Kobayashi:1973fv}. The neutrinos have tiny masses and mix between generations\cite{DoubleChooz:2011ymz,DayaBay:2012fng}. This indicates that lepton flavor is violated in the lepton sector. However, the SM cannot provide a reasonable explanation for either the small mass problem of neutrinos or the LFV. This suggests that the SM needs to be extended. Therefore the presence of any LFV signal may be regarded as evidence for the existence of NP beyond the SM\cite{Sun:2019wii}.
%The neutrinos have tiny masses and mix between generations\cite{DoubleChooz:2011ymz,DayaBay:2012fng}. If neutrinos possess the properties of Majorana particles, lepton number conservation is violated\cite{Bilenky:2005cp}. However, the lepton number in the SM  is conserved, which means that the LFV cannot be explained in the SM. Therefore the presence of any LFV signal may be regarded as evidence for the existence of NP beyond the SM\cite{Sun:2019wii}. 

In this work, we study the processes $Z\rightarrow l^{\pm}_{i}l^{\mp}_{j}$ in the B-LSSM\cite{OLeary:2011vlq,Abdallah:2014fra,Barger:2008wn,FileviezPerez:2008sx,FileviezPerez:2010ek}, where $B$ represents the baryon number and $L$ stands for the lepton number. The B-LSSM is the $U(1)$ extension of Minimal Supersymmetric Standard Model(MSSM)\cite{Nilles:1983ge,Haber:1984rc,Rosiek:1989rs}, whose local gauge group is $SU(3)_C\otimes{SU(2)_L}\otimes{U(1)_Y}\otimes{U(1)_{B-L}}$. The right-handed neutrinos are introduced in this model, the conservation of lepton number is broken by making the neutrinos acquire mass through the seesaw mechanism. In this model, the invariance under ${U(1)_{B-L}}$ gauge group imposes the R-parity conservation\cite{Aulakh:1999cd}.

These LFV processes $Z\rightarrow l^{\pm}_{i}l^{\mp}_{j}$ have been studied in many models\cite{Yang:2010iq,Han:2011aq,Zhang:2014osa,Dong:2017ksc,Wang:2022tgf}, and most of these are studied in the mass eigenstate basis. This method depends on the transposed matrix, and it is difficult to see the effect of the parameters in the analytical level. To this end, we use a method called the MIA\cite{Arganda:2015uca,Arganda:2016zvc,Wang:2022iaf,Moroi:1995yh} to calculate these processes, which works directly on the mass matrix in the electroweak interaction basis instead of dealing with the mass matrix after the diagonalization in the physical basis. Compared with the mass eigenstate basis, the MIA provides a set of simple analytic formulas for the form factors. It can be emphasized which parameters will be effectively tested in the future colliders. In addition, the MIA provides the model-independent parameterization methods, which can be easily applied to the extended model of MSSM. 

 The paper is organized as follows. In Sec.II, we summarize the B-LSSM briefly, including its superpotential, the general soft SUSY-breaking terms and needed mass matrices. In Sec.III, we give the analytical expressions for $Z\rightarrow l^{\pm}_{i}l^{\mp}_{j}$ in the mass eigenstate basis and electroweak interaction basis, respectively. In Sec.IV, we give the numerical analysis. The conclusion is discussed in Sec.V. The one-loop functions, the coupling coefficients in the mass eigenstate basis, the needed Feynman rules in the electroweak interaction basis, and the analytical expressions corresponding to Fig.\ref{Mia1} and Fig.\ref{Mia2} are collected in Appendix A, B, C and D, respectively.
 
 \section{The B-LSSM}
 
Compared with the MSSM, the B-LSSM adds two singlet Higgs superfields $\hat{\eta}$, $\hat{\bar{\eta}}$ and three generations of right-handed neutrinos superfields $\hat{\nu}_i^c$. The sneutrinos are disparted into CP-even sneutrinos and CP-odd sneutrinos, and their mass squared matrices are both extended to $6\times6$. 

In the TABLE \ref{B-Lquantum numbers}, we show the quantum numbers of gauge symmetry group for the chiral fields in the B-LSSM, the superpotential is given by
\begin{eqnarray}
&&W_{B-L} =Y_{u,ij}\hat{Q}_i\hat{H}_u\hat{U}_j^c-Y_{d,ij}\hat{Q}_i\hat{H}_d\hat{D}_j^c +Y_{e,ij}\hat{L}_i\hat{H}_d\hat{R}_j+\mu_{H}\hat{H}_d\hat{H}_u\nonumber\\&&~~~~~~~~ +Y_{x,ij}\hat{\nu}_i^c\hat{\eta}\hat{\nu}_j^c+Y_{\nu,ij}\hat{L}_i\hat{H}_u\hat{\nu}_j^c
-{\mu_{\eta}}\hat{\eta}\hat{\bar{\eta}},
\end{eqnarray}
where $i, j$ represent the generation indices, $Y_{u,ij}, Y_{d,ij}, Y_{e,ij}, Y_{x,ij}$ and $Y_{\nu,ij}$ correspond to the Yukawa coupling coefficients. $\mu_{H}$ and $\mu_{\eta}$ are both the parameters with mass dimension. $\mu_{H}$ indicates the supersymmetric mass between $SU(2)_{L}$ Higgs doublets $\hat{H}_d$ and $\hat{H}_u$, as well as $\mu_{\eta}$ represents the supersymmetric mass between $U(1)_{B-L}$ Higgs singlets $\hat{\eta}$ and $\hat{\bar{\eta}}$.

\begin{table}[t]
\caption{ \label{B-Lquantum numbers}  The chiral superfields and quantum numbers in the B-LSSM.}
\footnotesize
\begin{tabular}{|c|c|c|c|c|c|}
\hline
Superfield & Spin 0 & Spin \(\frac{1}{2}\) & Generations & \(U(1)_Y\otimes\, \text{SU}(2)_L\otimes\, \text{SU}(3)_C\otimes\, U(1)_{B-L}\) \\
\hline
\(\hat{H}_d\) & \(H_d\) & \(\tilde{H}_d\) & 1 & \((-\frac{1}{2},{\bf 2},{\bf 1},0) \) \\
\(\hat{H}_u\) & \(H_u\) & \(\tilde{H}_u\) & 1 & \((\frac{1}{2},{\bf 2},{\bf 1},0) \) \\
\(\hat{Q}_i\) & \(\tilde{Q}_i\) & \(Q_i\) & 3 & \((\frac{1}{6},{\bf 2},{\bf 3},\frac{1}{6}) \) \\
\(\hat{L}_i\) & \(\tilde{L}_i\) & \(L_i\) & 3 & \((-\frac{1}{2},{\bf 2},{\bf 1},-\frac{1}{2}) \) \\
\(\hat{D}_i^c\) & \(\tilde{D}_i^c\) & \(D_i^c\) & 3 & \((\frac{1}{3},{\bf 1},{\bf \overline{3}},-\frac{1}{6}) \) \\
\(\hat{U}_i^c\) & \(\tilde{U}_i^c\) & \(U_i^c\) & 3 & \((-\frac{2}{3},{\bf 1},{\bf \overline{3}},-\frac{1}{6}) \) \\
\(\hat{R}_i\) & \(\tilde{R}_i\) & \(R_i\) & 3 & \((1,{\bf 1},{\bf 1},\frac{1}{2}) \) \\
\(\hat{\nu}_i^c\) & \(\tilde{\nu}_i^c\) & \(\nu_i^c\) & 3 & \((0,{\bf 1},{\bf 1},\frac{1}{2}) \) \\
\(\hat{\eta}\) & \(\eta\) & \(\tilde{\eta}\) & 1 & \((0,{\bf 1},{\bf 1},-1) \) \\
\(\hat{\bar{\eta}}\) & \(\bar{\eta}\) & \(\tilde{\bar{\eta}}\) & 1 & \((0,{\bf 1},{\bf 1},1) \) \\
\hline
\end{tabular}
\end{table}

The $SU(2)_{L}\otimes U(1)_{Y}\otimes U(1)_{B-L}$ gauge group breaks to $U(1)_{em}$ as the Higgs fields obtain the vacuum expectation values (VEVs),
\begin{eqnarray}
&&H^{0}_{d}=\frac{1}{\sqrt{2}}(\phi_{d}+v_{d}+i\sigma_{d}),~ H^{0}_{u}=\frac{1}{\sqrt{2}}(\phi_{u}+v_{u}+i\sigma_{u}), \nonumber\\
&&\eta=\frac{1}{\sqrt{2}}(\phi_{\eta}+v_{\eta}+i\sigma_{\eta}),~ \bar{\eta}=\frac{1}{\sqrt{2}}(\phi_{\bar{\eta}}+v_{\bar{\eta}}+i\sigma_{\bar{\eta}}),
\end{eqnarray}
where $\sigma_{d}$, $\sigma_{u}$, $\sigma_{\eta}$, $\sigma_{\bar{\eta}}$ represent the CP-odd Higgs components and $\phi_{d}$, $\phi_{u}$, $\phi_{\eta}$, $\phi_{\bar{\eta}}$ correspond to the CP-even Higgs components. The VEVs of the Higgs singlets $\hat{\eta}$ and $\hat{\bar{\eta}}$ satisfy $u=\sqrt{v^{2}_{\eta}+v^{2}_{\bar{\eta}}}$. While the VEVs of the Higgs doublets $\hat{H}_d$ and $\hat{H}_u$ are $v_{d}$ and $v_{u}$, which satisfy $v=\sqrt{v^{2}_{d}+v^{2}_{u}}$. We take $\tan \beta^{\prime}=\frac{v_{\bar{\eta}}}{v_{\eta}}$ by analogy to the definition $\tan \beta=\frac{v_{u}}{v_{d}}$ in the MSSM.

The ${U(1)_{B-L}}$ group introduces a new gauge field: $B^{\prime BL}$. The coupling between the gauge fields $B^{\prime BL}$ and $B^{Y}$ is given by the matrix  $\left(\begin{array}{cc}
g_{Y} & g_{Y B}^{\prime} \\
g_{B Y}^{\prime} & g_{B-L}
\end{array}\right)$. With the condition of the two Abelian gauge groups unbroken, choosing matrix R in a proper form, one can write the coupling matrix as\cite{OLeary:2011vlq}
\begin{eqnarray}
\left(\begin{array}{cc}
g_{Y} & g_{Y B}^{\prime} \\
g_{B Y}^{\prime } & g_{B-L}
\end{array}\right) R^{T}=\left(\begin{array}{cc}
g_{1} & g_{Y B} \\
0 & g_{B}
\end{array}\right),
\end{eqnarray}
and the transformation of the corresponding gauge field is
\begin{eqnarray}
R\left(\begin{array}{c}
B^{ Y} \\
B^{\prime BL}
\end{array}\right)=\left(\begin{array}{c}
B \\
B^{\prime}
\end{array}\right).
\end{eqnarray}

The two Abelian groups in the B-LSSM produce a new effect called as the gauge kinetic mixing. Due to the presence of the kinetic mixing terms, the $B^{\prime}$ boson mixes at tree level
with the $B$ and $W^{3}$ bosons. Basis on ($B$, $W^{3}$, $B^{\prime}$), the mass matrix is
\begin{eqnarray}
\begin{pmatrix}
  g^{2}_{1}v^{2} & -g_{1}g_{2}v^{2} & g_{1}g_{YB}v^{2} \\
  -g_{1}g_{2}v^{2} & g^{2}_{2}v^{2} & -g_{2}g_{YB}v^{2} \\
  g_{1}g_{YB}v^{2} & -g_{2}g_{YB}v^{2} & g^{2}_{YB}v^{2}+g_{B}^{2}u^{2}
\end{pmatrix}.
\end{eqnarray}
This mass matrix can be diagonalized by a unitary matrix to get the physical mass eigenstates  $\gamma $, $Z$ and $Z^{\prime}$.
\begin{eqnarray}
\begin{pmatrix}
  B \\
  W \\
  B^{'}
  \end{pmatrix}=\begin{pmatrix}
                  \cos\theta_{W} & \cos\theta^{\prime}_{W}\sin\theta_{W} & -\sin\theta_{W}\sin\theta^{\prime}_{W} \\
                  \sin\theta_{W} & -\cos\theta_{W}\cos\theta^{\prime}_{W} & \cos\theta_{W}\sin\theta^{\prime}_{W} \\
                  0 & \sin\theta^{\prime}_{W} & \cos\theta^{\prime}_{W}
                \end{pmatrix}\begin{pmatrix}
                                 \gamma \\
                                 Z \\
                                 Z^{\prime}
                \end{pmatrix},
\end{eqnarray}
where $\theta_{W}$ and $\theta^{\prime}_{W}$ are the corresponding Weinberg angles.

The soft breaking terms in the B-LSSM are written as
\begin{eqnarray}
&&{\cal L}_{{soft}}^{B-L}=
- m_{\tilde{q},{i j}}^{2}\tilde{Q}_{{i}}^*\tilde{Q}_{{j}}- m_{\tilde{u},{i j}}^{2}\tilde{U}_{{i}}^*\tilde{U}_{{j}}- m_{\tilde{d},{i j}}^{2}(\tilde{D}^c_{{i}})^* \tilde{D}_{{j}}^c- m_{\tilde{L},{i j}}^{2}\tilde{L}_{{i}}^*\tilde{L}_{{j}}- m_{\tilde{E},{i j}}^{2}(\tilde{E}^c_{{i}})^* \tilde{E}_{{j}}^c\nonumber \\
&&\hspace{1.1cm}
-m_{{H}_d}^2 |{H}_d|^2-m_{{H}_u}^2 |{H}_u|^2-m_{{\eta}}^2 |{\eta}|^2-m_{{\bar\eta}}^2 |{\bar\eta}|^2
- m_{\tilde{\nu},{i j}}^{2}(\tilde{\nu}^c_{{i}})^* \tilde{\nu}_{{j}}^c+\Big[-B_{\mu}{H}_d {H}_u\nonumber \\&&\hspace{1.1cm}-B_{\eta}{\eta}{\bar\eta}
+T_{u}^{i j} \tilde{Q}_i \tilde{U}_j^c H_u+T_{d}^{i j} \tilde{Q}_i \tilde{D}_j^c H_d+T_{e}^{i j} \tilde{L}_i \tilde{E}_j^c H_u+T_{\nu}^{i j}H_u \tilde{\nu}^c_i \tilde{L}_j+T_{x}^{i j}{\eta} \tilde{\nu}^c_i\tilde{\nu}^c_j\nonumber \\
&&\hspace{1.1cm}-\frac{1}{2}({M}_1\lambda_{\tilde{B}}\lambda_{\tilde{B}} +{M}_2\lambda_{\tilde{W}}\lambda_{\tilde{W}}+{M}_3\lambda_{\tilde{g}}\lambda_{\tilde{g}}+2{M}_{B B'}\lambda_{\tilde{B}'}\lambda_{\tilde{B}}+{M}_{B'}\lambda_{\tilde{B}'}\lambda_{\tilde{B}'})+h.c.\Big],
\end{eqnarray}
where $\lambda_{\tilde{B}}, \lambda_{\tilde{W}}, \lambda_{\tilde{g}}$ and $\lambda_{\tilde{B}'}$ are the gauginos of $U(1)_Y, SU(2)_L, SU(3)_C$ and $U(1)_{B-L}$ respectively. For the soft breaking slepton mass matrices $m^{2}_{\tilde{L}, \tilde{E}}$ and the trilinear coupling matrix $T_{e}$, we introduce the slepton flavor mixings, which take into account the off-diagonal terms\cite{Zhang:2014osa,Calibbi:2017uvl}
\begin{eqnarray}
m_{\tilde{L}}^2\hspace{-0.1cm}=\hspace{-0.2cm}\left(\hspace{-0.1cm}\begin{array}{ccc}
m_{L}^2 & \delta_{12}^{LL}m_{LL}^2 &  \delta_{13}^{LL}m_{LL}^2\\
 \delta_{12}^{LL}m_{LL}^2 & m_{L}^2 & \delta_{23}^{LL}m_{LL}^2\\
 \delta_{13}^{LL}m_{LL}^2 & \delta_{23}^{LL}m_{LL}^2 &m_{L}^2
\end{array}\hspace{-0.1cm}\right)\hspace{-0.1cm},~~~
\end{eqnarray}
\begin{eqnarray}
m_{\tilde{E}}^2\hspace{-0.1cm}=\hspace{-0.2cm}\left(\hspace{-0.1cm}\begin{array}{ccc}
m_{E}^2 & \delta_{12}^{RR}m_{EE}^2 &  \delta_{13}^{RR}m_{EE}^2\\
 \delta_{12}^{RR}m_{EE}^2 & m_{E}^2 & \delta_{23}^{RR}m_{EE}^2\\
 \delta_{13}^{RR}m_{EE}^2 & \delta_{23}^{RR}m_{EE}^2 &m_{E}^2
\end{array}\hspace{-0.1cm}\right)\hspace{-0.1cm},~~~
\end{eqnarray}
\begin{eqnarray}
T_e\hspace{-0.1cm}=\hspace{-0.2cm}\left(\hspace{-0.1cm}\begin{array}{ccc}
1 & \delta_{12}^{LR} &  \delta_{13}^{LR}\\
 \delta_{12}^{LR} & 1 & \delta_{23}^{LR}\\
 \delta_{13}^{LR} & \delta_{23}^{LR} &1
\end{array}\hspace{-0.1cm}\right)\hspace{-0.1cm}A_e.
\end{eqnarray}

The mass matrices we used can be obtained by SARAH\cite{Staub:2015kfa}. We list some of these mass matrices here. In the B-LSSM, four 2-component spinors ($\lambda^{-}_{\tilde{W}}$, $\lambda^{+}_{\tilde{W}}$, $\tilde{H}^{-}$, $\tilde{H}^{+}$) form two four-component Dirac fermions (Charginos) $\chi^{+}$, $\chi^{-}$
 \begin{eqnarray}
M_{\chi^{\pm}}=\begin{pmatrix}
   M_{2} & \frac{1}{\sqrt{2}}g_{2}v_{u} \\
   \frac{1}{\sqrt{2}}g_{2}v_{d} & \mu_{H}
 \end{pmatrix}.
 \end{eqnarray}
This matrix is diagonalized by $U$ and $V$:
\begin{equation}
  U^{*} M_{\chi^{\pm}} V^{\dagger}=M_{\chi^{\pm}}^{\text {diag }}.
\end{equation}

Based on $\left(\lambda_{\tilde{B}}, \lambda^{3}_{\tilde{W}}, \tilde{H}_{d}^{0}, \tilde{H}_{u}^{0}, \lambda_{\tilde{B}^{\prime}}, \tilde{\eta}, \tilde{\bar{\eta}}\right)$, seven Majorana fermions(Neutralinos) can be obtained:
\begin{eqnarray}
M_{\chi^{0}}=\begin{pmatrix}
  M_{1} & 0 & -\frac{1}{2}g_{1}v_{d} & \frac{1}{2}g_{1}v_{u} & M_{BB^{'}} & 0 & 0 \\
  0 & M_{2} & \frac{1}{2}g_{2}v_{d} & -\frac{1}{2}g_{2}v_{u} & 0 & 0 & 0 \\
  -\frac{1}{2}g_{1}v_{d} & \frac{1}{2}g_{2}v_{d} & 0 & -\mu_{H} & -\frac{1}{2}g_{YB}v_{d} & 0 & 0 \\
  \frac{1}{2}g_{1}v_{u} & -\frac{1}{2}g_{2}v_{u} & -\mu_{H} & 0 & \frac{1}{2}g_{YB}v_{u} & 0 & 0 \\
  M_{BB^{'}} & 0 & -\frac{1}{2}g_{YB}v_{d} & \frac{1}{2}g_{YB}v_{u} & M_{B^{\prime}} & -g_{B}v_{\eta} & g_{B}v_{\bar{\eta}} \\
  0 & 0 & 0 & 0 & -g_{B}v_{\eta} & 0 & -\mu_{\eta} \\
  0 & 0 & 0 & 0 & g_{B}v_{\bar{\eta}} & -\mu_{\eta} & 0
\end{pmatrix}.
\end{eqnarray}
This matrix is diagonalized by $N$:
\begin{equation}
  N^{*} M_{\chi^{0}} N^{\dagger}=M_{\chi^{0}}^{\text {diag }}.
\end{equation}

Based on $(\tilde{L}^{I}_{2}, \tilde{R})$, the mass squared matrix for sleptons reads
\begin{eqnarray}
&&\tilde{M}^{2}_{\tilde{l}}=\begin{pmatrix}
                              (M^{2}_{L})_{LL} & (M^{2}_{L})_{LR} \\
                              (M^{2}_{L})^{\dagger}_{LR} & (M^{2}_{L})_{RR}
                            \end{pmatrix},
\end{eqnarray}
\begin{eqnarray}
&&(M^{2}_{L})_{LL}=m^{2}_{\tilde{L}}+\mathbf{1}\frac{1}{8}\big((g^{2}_{1}+g^{2}_{YB})(-v^{2}_{u}+v^{2}_{d})+g^{2}_{2}(-v^{2}_{d}+v^{2}_{u})+g_{YB}g_{B}(-2v^{2}_{\bar{\eta}}+2v^{2}_{\eta}-v^{2}_{d}+v^{2}_{u}) \nonumber\\
&&\;\;\;\;\;\;\;\;\;\;\;\;\;\;\;\;\; +2g^{2}_{B}(-v^{2}_{\bar{\eta}}+v^{2}_{\eta})\big)+\frac{1}{2}v^{2}_{d}Y^{\dagger}_{e}Y_{e}, \nonumber\\
&&(M^{2}_{L})_{RR}=m^{2}_{\tilde{E}}+\mathbf{1}\frac{1}{8}\big(2(g^{2}_{1}+g^{2}_{YB})(-v^{2}_{d}+v^{2}_{u})-2g^{2}_{B}(-v^{2}_{\bar{\eta}}+v^{2}_{\eta})+g_{B}g_{YB}(4v^{2}_{\bar{\eta}}-4v^{2}_{\eta}-v^{2}_{d}+v^{2}_{u})\big) \nonumber\\
&&\;\;\;\;\;\;\;\;\;\;\;\;\;\;\;\;\;  +\frac{1}{2}v^{2}_{d}Y_{e}Y^{\dagger}_{e}, \nonumber\\
&&(M^{2}_{L})_{LR}=\frac{1}{\sqrt{2}}(v_{d}T_{e}^{\dagger}-v_{u}\mu Y^{\dagger}_{e}).
\label{eq111}
\end{eqnarray}
This matrix is diagonalized by $Z^{E}$:\begin{eqnarray}
 &&Z^{E} \tilde{M}^{2}_{\tilde{l}} Z^{E,\dagger}=({M^{2}_{\tilde{l}}})^{\text {diag }}.
\end{eqnarray}

 \section{The lfv decays $Z\rightarrow l^{\pm}_{i}l^{\mp}_{j}$}
The corresponding effective amplitude for $Z \rightarrow l^{\pm}_{i}l^{\mp}_{j}$, obtained through the effective Lagrangian, can be written as follows\cite{Flores-Tlalpa:2001vbz}:
 \begin{eqnarray}
\mathcal{M}\left(Z \rightarrow l_{i}^{ \pm} l_{j}^{\mp}\right)=\bar{l}_{i} \gamma_{\mu}\left(A_{L} P_{L}+A_{R} P_{R}\right) l_{j} Z^{\mu}.
\label{Brz}
 \end{eqnarray}
Then, we can obtain the branching ratios of $Z\rightarrow l^{\pm}_{i}l^{\mp}_{j}$:
\begin{eqnarray}
  Br(Z\rightarrow l^{\pm}_{i}l^{\mp}_{j})=\frac{1}{12\pi}\frac{m_{Z}}{\Gamma_{Z}}(|A_{L}|^{2}+|A_{R}|^{2}),
\end{eqnarray}
where $\Gamma_{Z}$ denotes the total decay width of $Z$ boson. In the numerical calculation, we choose $\Gamma_{Z}\simeq 2.4952$GeV\cite{ParticleDataGroup:2022pth}. The coefficients $A_{L,R}$ can be obtained from the amplitudes of the Feynman diagrams.
 \subsection{The coefficients in the mass eigenstate basis}

\begin{figure}[t]
  \centering
  \includegraphics[width=10cm]{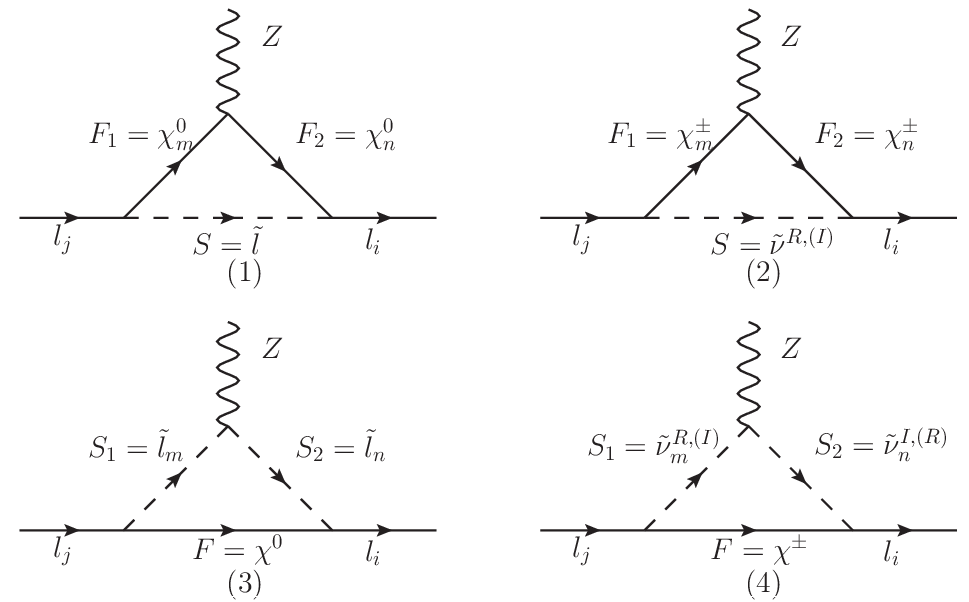}
  \caption{Feynman diagrams for $Z\rightarrow l^{\pm}_{i}l^{\mp}_{j}$ in the mass eigenstate basis.}
  \label{mass}
\end{figure}

 The corresponding Feynman diagrams for $Z\rightarrow l^{\pm}_{i}l^{\mp}_{j}$  are shown in Fig.\ref{mass}, where F and S represent Dirac (Majorana) fermions and scalar bosons, respectively. We then give specific expressions for the coefficients of Eq.(\ref{Brz}) \begin{eqnarray}
&&A_{L}=\frac{1}{2}\sum_{F= \chi^{\pm}, \chi^{0}}\sum_{S=\tilde{\nu}^{R,I}, \tilde{l}}\big[\frac{2m_{F_{1}}m_{F_{2}}}{\Lambda^{2}}H^{{SF_{2}l_{i}}}_{R}H^{{ZF_{1}F_{2}}}_{L}H^{{S^{\ast}F_{1}l_{j}}}_{L}G_{1}(x_{S},x_{F_{1}},x_{F_{2}}) \nonumber\\
&&~~~~~~~-H^{{SF_{2}l_{i}}}_{R}H^{{ZF_{2}F_{1}}}_{L}H^{{S^{\ast}F_{1}l_{j}}}_{L}G_{2}(x_{S},x_{F_{1}},x_{F_{2}})\big]  \nonumber\\
&&~~~~~~~-\frac{1}{2}\sum_{F= \chi^{\pm}, \chi^{0}}\sum_{S=\tilde{\nu}^{R,I}, \tilde{l}}\big[H^{{FS_{2}l_{i}}}_{R}H^{{ZS^{\ast}_{2}S_{1}}}_{L}H^{{S_{1}^{\ast}Fl_{j}}}_{L}G_{2}(x_{F},x_{S_{1}},x_{S_{2}})\big], \nonumber\\
&&A_{R}=A_{L}|_{L\leftrightarrow R}.
\end{eqnarray}
Here, the concrete expressions for coupling coefficients $H_{L,R}$ and the one-loop functions can be found in Appendix \ref{OLF}. Besides, we take $x=m^{2}/\Lambda^{2}$ with $m$ being the mass of the corresponding particle, and $\Lambda$ representing the energy scale of the NP.

 \subsection{The analytical results obtained through the MIA method}
In contradistinction to the mass eigenstate basis, the Feynman diagrams of the MIA postulate that the external particles are assumed in the mass basis, whereas the internal sparticles operate within the electroweak interaction basis. The diagonal terms of the mass matrix are defined as the sparticle mass, and the off-diagonal terms are defined as the interaction vertices considered as mass insertions\cite{Dedes:2015twa,Rosiek:2015jua}. The mass insertions can be categorized into two types: the flavor preservation and the flavor violation. The LFV comes from the off-diagonal terms of the sleptons (sneutrinos) mass matrix. Thus, we define $(i\ne j)$\cite{Calibbi:2017uvl}
\begin{eqnarray}
  \Delta^{LL}_{ij}=(M^{2}_{L})^{ij}_{LL}=(M^{2}_{\tilde{\nu}_{L}^{R,I}})^{ij}_{LL},~\Delta^{LR}_{ij}=(M^{2}_{L})^{ij}_{LR},~\Delta^{RR}_{ij}=(M^{2}_{L})^{ij}_{RR},
\end{eqnarray}
The terms $L$ and $R$ in $\Delta^{AB}_{ij}(A,B=L,R)$ specifically denote left-handed sleptons (sneutrinos) and right-handed sleptons. After our analysis, only these mass insertions change the flavor.

For the insertion of $\Delta^{LR}_{ii}$, this case is considered an insertion that does not change flavor. Therefore, it is necessary to introduce an additional insert $\Delta^{LL}_{ij}$ or $\Delta^{RR}_{ij} (i\ne j)$ to change the flavor to meet the requirements. 

The LFV processes in the MIA need to consider the trilinear couplings under the interaction eigenstate, so we show some couplings needed in this work as follows. The lepton-charginos-CP-even(odd) sneutrinos are deduced as:
\begin{eqnarray}
&&{\cal L}_{\bar{l}_j\chi^-\tilde{\nu}^R}=\frac{i}{\sqrt{2}}\bar{l}_j
\tilde{\nu}_L^R[Y_l^jP_L\tilde{H}^-+g_2P_R\lambda^{-}_{\tilde{W}}],\nonumber \\&&
{\cal L}_{\bar{l}_j\chi^-\tilde{\nu}^I}=\frac{1}{\sqrt{2}}\bar{l}_j
\tilde{\nu}_L^I[-Y_l^jP_L\tilde{H}^-+g_2P_R\lambda^{-}_{\tilde{W}}].
\end{eqnarray}
The lepton-neutralinos-sleptons are deduced as:
\begin{eqnarray}
&&{\cal L}_{\bar{l}_j\chi^0\tilde{l}}=i\bar{l}_j\Big\{-\Big[\frac{1}{\sqrt{2}}
\Big(2g_{1}P_{L}\lambda_{\tilde{B}}+(g_B+2g_{YB})P_{L}\lambda_{\tilde{B}'}\Big)\tilde{R}+Y_l^{j}P_{L}\tilde{H}^0_d\tilde{L}\Big]\nonumber \\&&\hspace{2.3cm}+\Big[\frac{1}{\sqrt{2}}
\Big(g_{2}P_{R}\lambda^{3}_{\tilde{W}}+g_{1}P_{R}\lambda_{\tilde{B}}+(g_B+g_{YB})P_{R}\lambda_{\tilde{B}'}\Big)\tilde{L}-Y_l^{j}P_{R}\tilde{H}^0_d\tilde{R}\Big]\Big\}.
\end{eqnarray}

\begin{figure}
  \centering
  \includegraphics{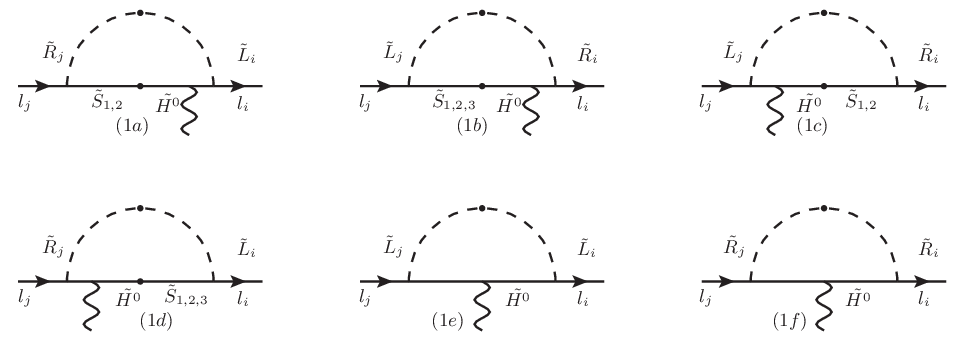}
  \includegraphics{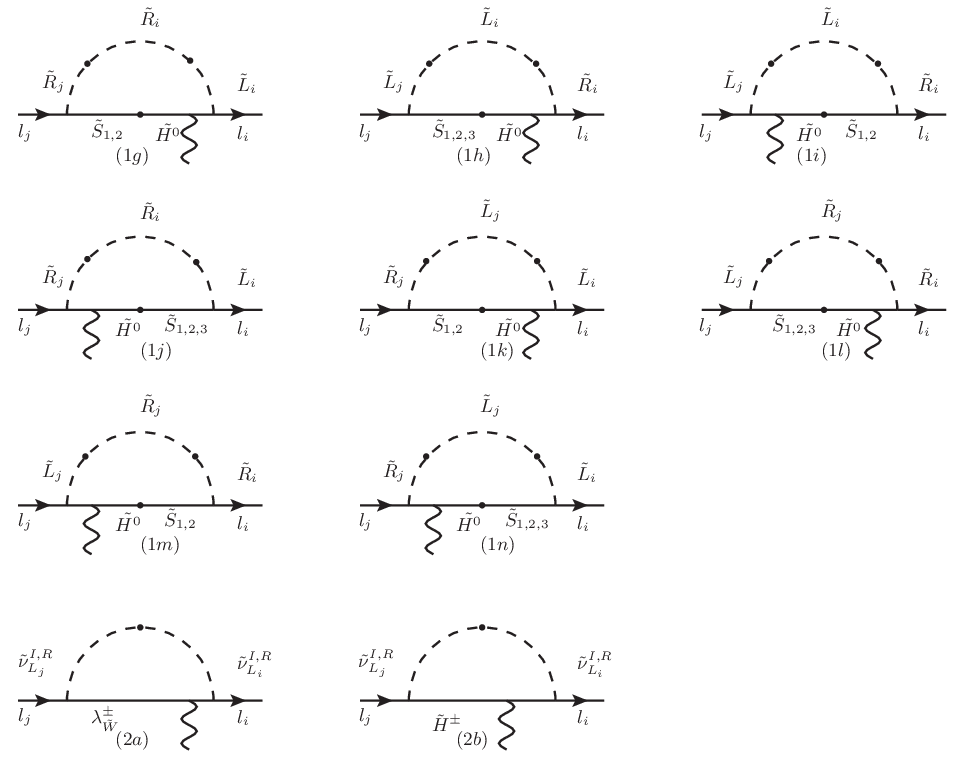}
  \caption{The Feynman diagrams for $Z\rightarrow l^{\pm}_{i}l^{\mp}_{j}$  in the electroweak interaction basis.}
  \label{Mia1}
\end{figure}

\begin{figure}
  \centering
  \includegraphics{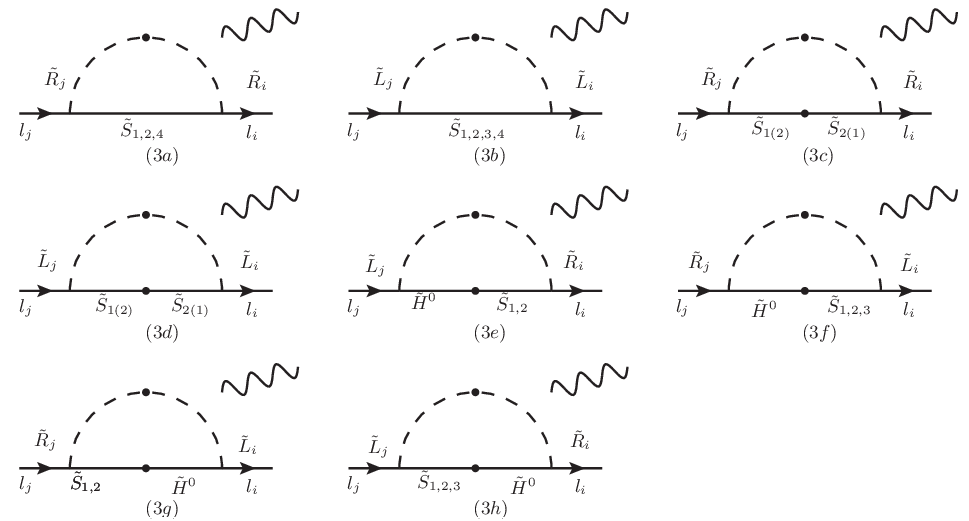}
  \includegraphics{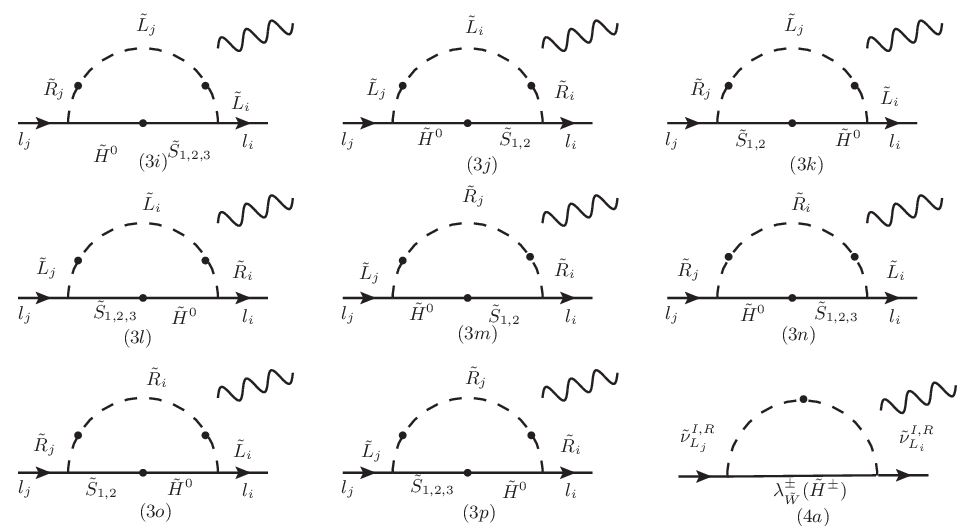}
  \caption{The Feynman diagrams for $Z\rightarrow l^{\pm}_{i}l^{\mp}_{j}$  in the electroweak interaction basis.}
  \label{Mia2}
\end{figure}

We show Feynman diagrams using the MIA in Fig.\ref{Mia1} and Fig.\ref{Mia2}. Here, $\tilde{S}_{1,2,3,4}$ is defined as $\lambda_{\tilde{B}}$, $\lambda_{\tilde{B}^{\prime}}$, $\lambda^{3}_{\tilde{W}}$, $\tilde{H}^{0}$. $m_{l_{j}}$ is the j-th generation lepton mass. $\tilde{R}_{i}$ is the i-th generation right-handed slepton, $\tilde{L}_{i}$ is the i-th generation left-handed slepton, $\tilde{\nu}^{R,I}_{L_{j}}$ is the j-th generation left-handed CP-even(odd) sneutrino. The Feynman diagrams including the right-handed sneutrino are strongly suppressed by the coupling parameter $Y_{\nu}$, and the situations with right-handed sneutrino are neglected. To save space, we present the contributions from several special Feynman diagrams, the rest of which can be found in the Appendix \ref{MR}. The one-loop contributions from Fig.\ref{Mia1}(2a) are shown as:
\begin{eqnarray}
&&A_{L}(2a)=-\frac{1}{4\Lambda ^{2}}g^{3}_{2}\cos \theta \cos \theta^{\prime} \Delta_{ij}^{LL}[G_{3}(x_{2},x_{\tilde{\nu}^{I}_{L_{j}}},x_{\tilde{\nu}^{I}_{L_{i}}})-2x_{2}I_{2}(x_{2},x_{\tilde{\nu}^{I}_{L_{j}}},x_{\tilde{\nu}^{I}_{L_{i}}})]\nonumber\\
&&~~~~~~~~~~~~~-\frac{1}{4\Lambda ^{2}}g^{3}_{2}\cos \theta \cos \theta^{\prime} \Delta_{ij}^{LL}[G_{3}(x_{2},x_{\tilde{\nu}^{R}_{L_{j}}},x_{\tilde{\nu}^{R}_{L_{i}}})-2x_{2}I_{2}(x_{2},x_{\tilde{\nu}^{R}_{L_{j}}},x_{\tilde{\nu}^{R}_{L_{i}}})].
\label{eq21}
\end{eqnarray}
The one-loop contributions from Fig.\ref{Mia1}(2b) are shown as:
\begin{eqnarray}
&&A_{R}(2b)=\frac{1}{2\Lambda ^{2}}Y_{e,jj}Y_{e,ii}(-g_{1}\sin \theta \cos \theta^{\prime}+g_{2}\cos \theta \cos \theta^{\prime}+g_{YB}\sin \theta^{\prime}) \nonumber\\
&&~~~~~~~~~~~~\times \Delta_{ij}^{LL}[G_{3}(x_{\mu_{H}},x_{\tilde{\nu}^{I}_{L_{j}}},x_{\tilde{\nu}^{I}_{L_{i}}})-2x_{2}I_{2}(x_{\mu_{H}},x_{\tilde{\nu}^{I}_{L_{j}}},x_{\tilde{\nu}^{I}_{L_{i}}})] \nonumber\\
&&~~~~~~~~~~~~+\frac{1}{2\Lambda ^{2}}Y_{e,jj}Y_{e,ii}(-g_{1}\sin \theta \cos \theta^{\prime}+g_{2}\cos \theta \cos \theta^{\prime}+g_{YB}\sin \theta^{\prime}) \nonumber\\
&&~~~~~~~~~~~~\times \Delta_{ij}^{LL}[G_{3}(x_{\mu_{H}},x_{\tilde{\nu}^{R}_{L_{j}}},x_{\tilde{\nu}^{R}_{L_{i}}})-2x_{2}I_{2}(x_{\mu_{H}},x_{\tilde{\nu}^{R}_{L_{j}}},x_{\tilde{\nu}^{R}_{L_{i}}})].
\label{eq22}
\end{eqnarray}
The one-loop contributions from Fig.\ref{Mia1}(4a) are shown as:
\begin{eqnarray}
&&A_{R}(4a)=-\frac{1}{8\Lambda ^{2}}Y_{e,jj}Y_{e,ii}\big(g_{2}\cos \theta \cos \theta^{\prime}+g_{2}\sin \theta \cos \theta^{\prime}-(g_{B}+g_{YB})\sin \theta^{\prime}\big) \nonumber\\
&&~~~~~~~~~~~~\times \Delta_{ij}^{LL}[G_{3}(x_{\tilde{\nu}^{I}_{L_{j}}},x_{\tilde{\nu}^{I}_{L_{i}}},x_{\mu_{H}})+G_{3}(x_{\tilde{\nu}^{I}_{L_{i}}},x_{\tilde{\nu}^{I}_{L_{j}}},x_{\mu_{H}})] \nonumber\\
&&~~~~~~~~~~~~-\frac{1}{8\Lambda ^{2}}Y_{e,jj}Y_{e,ii}\big(g_{2}\cos \theta \cos \theta^{\prime}+g_{2}\sin \theta \cos \theta^{\prime}-(g_{B}+g_{YB})\sin \theta^{\prime}\big) \nonumber\\
&&~~~~~~~~~~~~\times \Delta_{ij}^{LL}[G_{3}(x_{\tilde{\nu}^{R}_{L_{j}}},x_{\tilde{\nu}^{R}_{L_{i}}},x_{\mu_{H}})+G_{3}(x_{\tilde{\nu}^{R}_{L_{i}}},x_{\tilde{\nu}^{R}_{L_{j}}},x_{\mu_{H}})],\\
&&A_{L}(4a)=-\frac{1}{8\Lambda ^{2}}g^{2}_{2}\big(g_{2}\cos \theta \cos \theta^{\prime}+g_{1}\sin \theta \cos \theta^{\prime}-(g_{B}+g_{YB})\sin \theta^{\prime}\big) \nonumber\\
&&~~~~~~~~~~~~\times \Delta_{ij}^{LL}[G_{3}(x_{\tilde{\nu}^{I}_{L_{j}}},x_{\tilde{\nu}^{I}_{L_{i}}},x_{2})+G_{3}(x_{\tilde{\nu}^{I}_{L_{i}}},x_{\tilde{\nu}^{I}_{L_{j}}},x_{2})] \nonumber\\
&&~~~~~~~~~~~~-\frac{1}{8\Lambda ^{2}}g^{2}_{2}\big(g_{2}\cos \theta \cos \theta^{\prime}+g_{1}\sin \theta \cos \theta^{\prime}-(g_{B}+g_{YB})\sin \theta^{\prime}\big) \nonumber\\
&&~~~~~~~~~~~~\times \Delta_{ij}^{LL}[G_{3}(x_{\tilde{\nu}^{R}_{L_{j}}},x_{\tilde{\nu}^{R}_{L_{i}}},x_{2})+G_{3}(x_{\tilde{\nu}^{R}_{L_{i}}},x_{\tilde{\nu}^{R}_{L_{j}}},x_{2})].
\label{eq41}
\end{eqnarray}
In these equations, the contributions from the parameters to the LFV process can be clearly seen. Especially, we can find that the contributions coressponding to Eq.(\ref{eq21})-Eq.(\ref{eq41}) depend on the parameter $\Delta_{ij}^{LL}$. Besides, from the equations listed in Appendix \ref{MR}, it follows that the one-loop contributions from Fig.\ref{Mia1} and Fig.\ref{Mia2} are affected by $\Delta^{AB}_{ij}(A,B=L,R)$.  Consider the limiting case: $\Delta^{AB}_{ij}(A,B=L,R)=0$, then $Br(Z\rightarrow l^{\pm}_{i}l^{\mp}_{j})=0$.

In order to more intuitively analyze the factors that affect LFV processes $Z\rightarrow l^{\pm}_{i}l^{\mp}_{j}$, we suppose that the sparticle masses are degenerate.  In other words, we give the one-loop results in the extreme case where the sparticle masses and mass insertion treams are equal to $\Lambda$:
\begin{eqnarray}
&&M_{1}=M_{2}=\mu_{H}=M_{B^{\prime}}=m_{L}=m_{E}=\tilde{M}_{\tilde{l},\tilde{\nu},ii}=\Lambda,\nonumber\\
&&M_{BB^{\prime}}=m_{LL}=m_{EE}=A_{e}=\Lambda.
\end{eqnarray}
In this extreme degeneracy case, the corresponding one-loop functions in the electroweak interaction basis are subjected to limit operations, which are equal to some specific values:\begin{eqnarray}
&&G_{3}(1, 1, 1)=\frac{1}{48\pi^{2}},~~~I_{2}(1, 1, 1)=-\frac{1}{96\pi^{2}}, \nonumber\\
&&G_{4}(1, 1, 1, 1)=\frac{1}{192\pi^{2}},~~~G_{5}(1, 1, 1, 1)=-\frac{1}{192\pi^{2}}.
\end{eqnarray}
And mass insertions $\Delta^{AB}_{ij}(A, B=L, R)$ that change leptons flavor become the product of $\Lambda^{2}$ and the dimensionless parameters $\delta^{AB}_{ij}(A, B=L, R)$:
\begin{eqnarray}
&&\Delta^{LL}_{ij}\approx \Lambda^{2} \delta^{LL}_{ij},~~~\Delta^{LR}_{ij}\approx \Lambda^{2} \delta^{LR}_{ij},~~~\Delta^{RR}_{ij}\approx \Lambda^{2} \delta^{RR}_{ij}.
\end{eqnarray}
 Then, we use these results to simplify all obtained coefficients to get some approximate results:\begin{eqnarray}
&&A_{R}^{\delta_{ij}^{RR}}\approx\frac{1}{48\pi^{2}}C_{R}[g_{1}^{2}+(g_{B}+2g_{YB})^{2}], \nonumber\\
&&A_{L}^{\delta_{ij}^{RR}}\approx\frac{1}{48\pi^{2}}C_{R}g_{1}(g_{B}+2g_{YB}), \nonumber\\
&&A_{L}^{\delta_{ij}^{LL}}\approx-\frac{1}{12\pi^{2}}C_{L}[g_{1}^{2}+(g_{B}+g_{YB})^{2}+g^{2}_{2}]-\frac{1}{384\pi^{2}}C_{L}g_{1}(g_{B}+g_{YB}) \nonumber\\
&&~~~~~~~~~~-\frac{1}{24\pi^{2}}g_{2}^{2}[g_{2}\cos \theta \cos \theta^{\prime}+g_{1}\sin \theta \cos \theta^{\prime}-(g_{B}+g_{YB})\sin \theta^{\prime}]\nonumber\\
&&~~~~~~~~~~-\frac{1}{48\pi^{2}}g_{2}^{3}\cos \theta \cos \theta^{\prime}, \nonumber\\
&&A_{R}^{\delta_{ij}^{LR}}\approx \frac{1}{192\pi^{2}}\times\frac{m_{l_{i}}+m_{l_{j}}}{8\Lambda}\tan \beta(C_{L}-C_{R}-2E)[2g_{1}^{2}+g_{YB}(g_{B}+2g_{YB})], \nonumber\\
&&A_{L}^{\delta_{ij}^{LR}}\approx \frac{1}{192\pi^{2}}\times\frac{m_{l_{i}}+m_{l_{j}}}{8\Lambda}\tan \beta(C_{R}-C_{L}-2E)[g_{1}^{2}+g_{YB}(g_{B}+2g_{YB})-g_{2}^{2}].
\label{eqsdelta}
\end{eqnarray}
Here, the coefficients $C_{L}$, $C_{R}$ and $E$ are collected in the Appendix \ref{MR}. By looking at these equations, we see that $\frac{m_{l_{i}}+m_{l_{j}}}{\Lambda}\ll1$ in  \textbf{$A_{L, R}^{\delta_{ij}^{LR}}$}, which lead to the numerical results of \textbf{$A_{L, R}^{\delta_{ij}^{LR}}$} being much smaller than that of $A_{L, R}^{\delta_{ij}^{RR}}$ and $A_{L}^{\delta_{ij}^{LL}}$.
This point will be confirmed in the numerical results of \textrm{Section IV}. In addition, we also see  that parameter \textbf{$\tan \beta$} is only coupled with $A_{L, R}^{\delta_{ij}^{LR}}$ in Eqs.(\ref{eqsdelta}), and the numerical results of \textbf{$A_{L, R}^{\delta_{ij}^{LR}}$} analyzed above are very small, so the influence of \textbf{$\tan \beta$} on $Z\rightarrow l^{\pm}_{i}l^{\mp}_{j}$ can be ignored.

\section{The numerical analysis}
In this section, we give the numerical analysis. In the B-LSSM, the $(g-2)_{\mu}$ and $l^{-}_{j} \rightarrow l^{-}_{i} \gamma$ have been discussed in Refs\cite{Yang:2018guw,Dong:2024lvs}, these two papers discuss the influence of sensitive parameters on $(g-2)_{\mu}$ and $l^{-}_{j} \rightarrow l^{-}_{i} \gamma$ under the mass eigenstates basis and the electroweak interaction basis.  Consider the constraints of $(g-2)_{\mu}$ and $l^{-}_{j} \rightarrow l^{-}_{i} \gamma$, we restudy the Z boson decays $Z\rightarrow l^{\pm}_{i}l^{\mp}_{j}$.

 The updated experimental data on the mass of $Z^{\prime}$ boson indicates $M^{\prime}_{Z}>5.1$ TeV with $95\%$ confidence level (C.L.)\cite{ATLAS:2019erb}, we choose $M^{\prime}_{Z}=5.4$ TeV in the following. Refs \cite{Cacciapaglia:2006pk,Carena:2004xs} give an upper bound on the ratio between the $Z^{\prime}$ mass and its gauge coupling at $99\%$ C.L. As $M^{\prime}_{Z}/g_{B}\geq 6$ TeV, the scope of $g_{B}$ is $0<g_{B}\lesssim0.9$. The LHC  experimental data constrains $\tan \beta^{\prime} <1.5$\cite{Basso:2015xna}. The coupling parameter $g_{YB}$ will be taken around $-0.45 <g_{YB}<-0.05$\cite{OLeary:2011vlq}. The large $\tan \beta$ has been excluded by the $\bar{B}\rightarrow X_{s}\gamma$ experiment\cite{Mahmoudi:2007gd,Olive:2008vv}. We select the suitable parameters, which are chosen as below:
\begin{eqnarray}
&&M_{1}=0.6 \mathrm{TeV}, A_{e}=0.5 \mathrm{TeV}, \Lambda=0.95 \mathrm{TeV}, M_{2}=0.7 \mathrm{TeV}, M_{BB^{\prime}}=0.8 \mathrm{TeV}, \nonumber\\
&&g_{YB}=-0.3, g_{B}=0.6, \tan \beta=20, \tan \beta^{\prime}=1.15, \mu_{H}=0.85 \mathrm{TeV},\nonumber\\
&&m_{L}=m_{E}=1.2 \mathrm{TeV}, m_{EE}=m_{LL}=1 \mathrm{TeV}.
\end{eqnarray}
\subsection{Constraints on $(g-2)_{\mu}$ and $l^{-}_{j} \rightarrow l^{-}_{i} \gamma$}
 \begin{figure}[t]
  \centering
  \includegraphics[width=5.4cm]{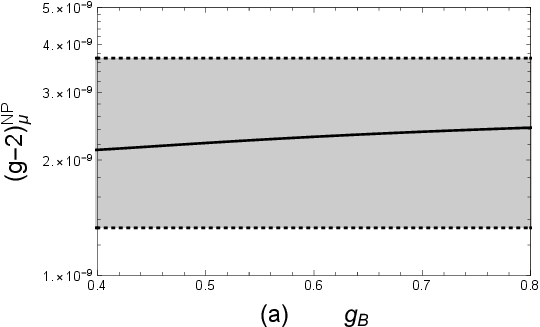}
  \includegraphics[width=5.4cm]{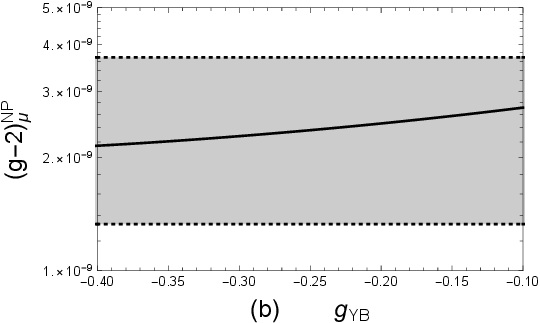}
  \includegraphics[width=5.4cm]{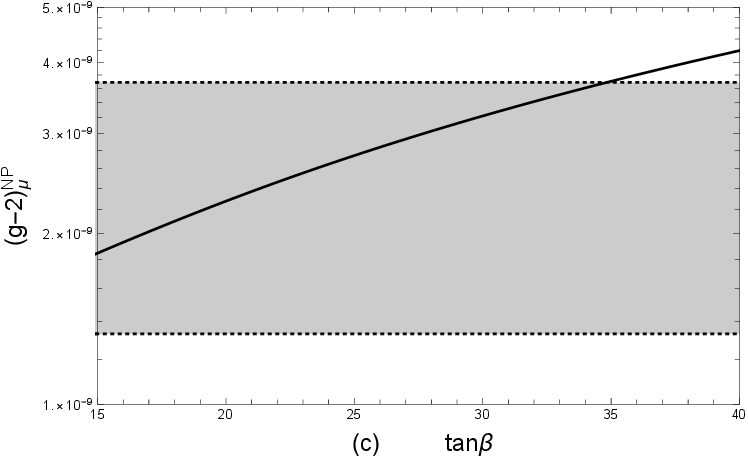}
  \includegraphics[width=5.4cm]{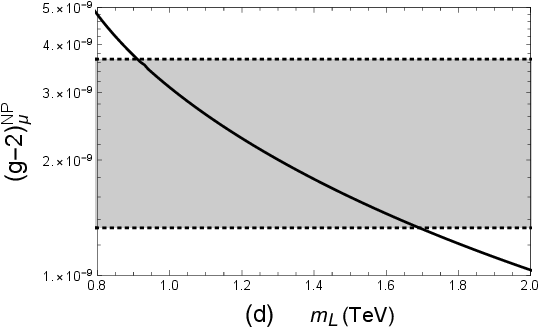}
  \includegraphics[width=5.4cm]{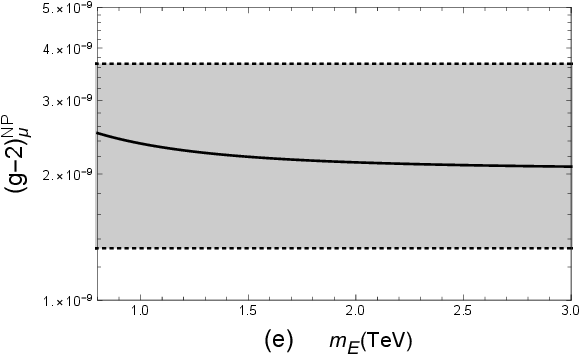}
  \includegraphics[width=5.4cm]{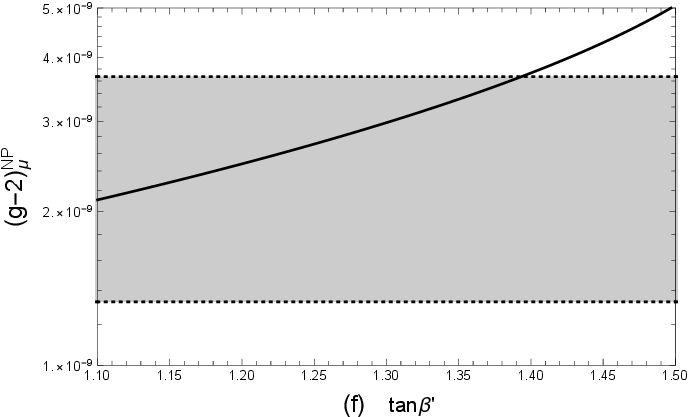}
  \caption{$\Delta a^{NP}_{\mu}$ versus $g_{B}$, $g_{YB}$, $\tan \beta$, $m_{L}$, $m_{E}$ and $\tan \beta^{\prime}$, the gray area denotes the experimental $2\sigma$ interval.}
\label{g-2}
\end{figure}
The new experimental average for the difference between the experimental measurement and SM theoretical prediction
of $(g-2)_{\mu}$ is given by\cite{Muong-2:2021ojo}
\begin{eqnarray}
  \Delta a_{\mu}=a^{exp}_{\mu}-a^{SM}_{\mu}=(25.1\pm5.9)\times10^{-10}.
\end{eqnarray}
In Fig.\ref{g-2}, we show $\Delta a^{NP}_{\mu}$ versus $g_{B}$, $g_{YB}$, $\tan \beta$, $m_{L}$, $m_{E}$ and $\tan\beta^{\prime}$. Under the $2\sigma$ limitation, the parameters $g_{B}$, $g_{YB}$ and $m_{E}$ are not sensitive to change. In addition, $m_{L}$, $\tan\beta$ and $\tan\beta^{\prime}$ obtain reasonable parameter spaces, which are constrained as: $\tan\beta<35$, $0.9\mathrm{TeV}<m_{L}<1.7 \mathrm{TeV}$, $\tan \beta^{\prime}<1.4$.

After considering the constraints of $(g-2)_{\mu}$, we proceed to analyze the process of $\mu \rightarrow e \gamma$. The latest experimental datas for the CLFV processes $l^{-}_{j} \rightarrow l^{-}_{i} \gamma$ at $90\%$ C.L. are \cite{MEG:2016leq,Belle:2021ysv}
 \begin{eqnarray}
   Br(\mu \rightarrow e \gamma)<4.2 \times 10^{-13}, Br(\tau \rightarrow e \gamma)<5.6 \times 10^{-8}, Br(\tau \rightarrow \mu \gamma)<4.2 \times 10^{-8}.
 \end{eqnarray}
 \begin{figure}[t]
  \centering
  \centering
  \includegraphics[width=5.4cm]{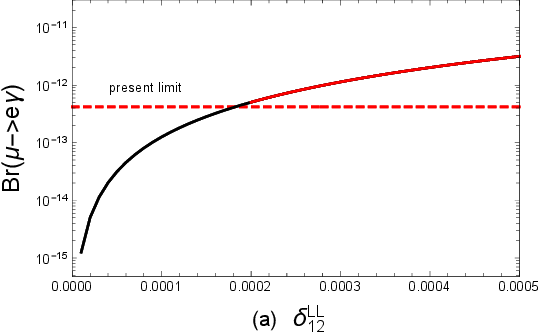}
  \includegraphics[width=5.4cm]{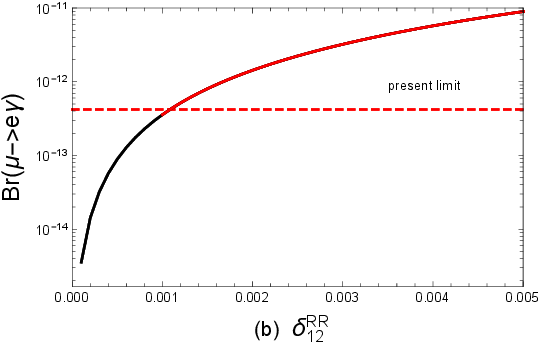}
  \includegraphics[width=5.4cm]{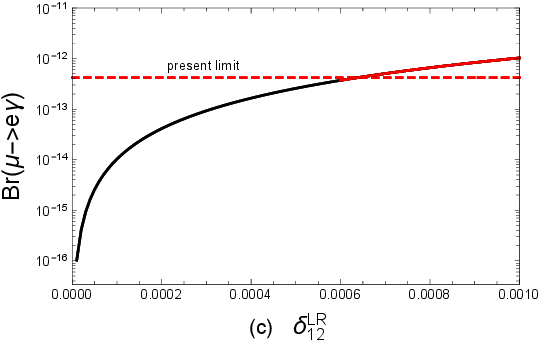}
  \caption{The LFV rates for $\mu \rightarrow e \gamma$ versus $\delta^{AB}_{ij}$, where the dashed red line denotes the present limit.}
\label{llrdelta}
\end{figure}
\begin{figure}[t]
  \centering
  \includegraphics[width=5.7cm]{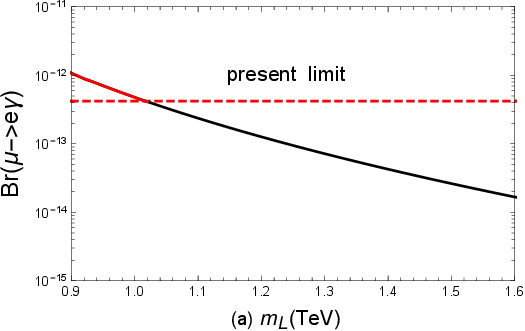}
  \includegraphics[width=5.7cm]{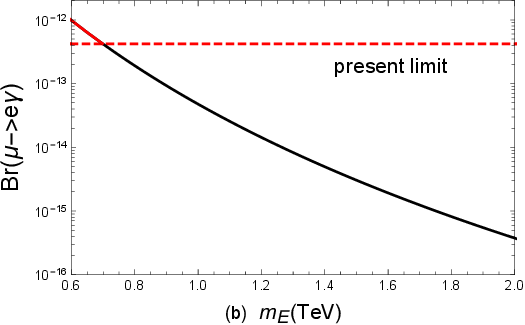}
  \includegraphics[width=5.7cm]{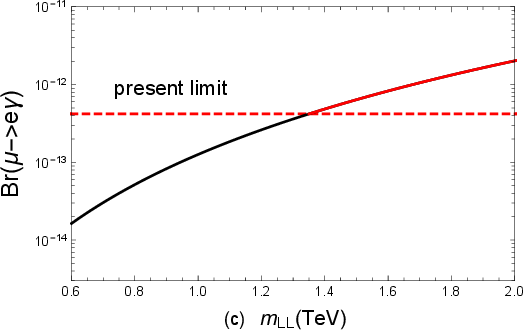}
  \includegraphics[width=5.7cm]{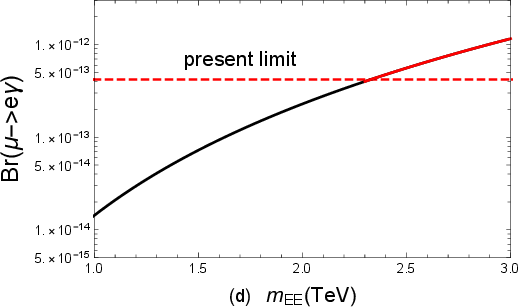}
  \caption{The LFV rates for $\mu \rightarrow e \gamma$ versus $M_{L}$, $M_{E}$, $M_{LL}$ and $M_{EE}$, where the dashed red line denotes the present limit.}
\label{llrm}
\end{figure}

 In Fig.\ref{llrdelta}, we plot the LFV rates for $l^{-}_{j} \rightarrow l^{-}_{i} \gamma$ versus $\delta^{AB}_{ij}(A,B=L,R)$. The branching ratios increase with the increase of the parameters $\delta^{AB}_{ij}$ and reach the present limit. It can be seen that the parameters are constrained by the present limit:
\begin{eqnarray}
&&\delta^{LL}_{12}<0.0002,~~~\delta^{RR}_{12}<0.001,~~~\delta^{LR}_{12}<0.0006.
\end{eqnarray}
In Fig.\ref{llrm}, we study the LFV rates for $\mu \rightarrow e \gamma$ versus $m_{L}$, $m_{E}$, $m_{LL}$ and $m_{EE}$. In Fig.\ref{llrm} (a) and (c),  we set $\delta^{RR}_{12}=\delta^{LR}_{12}=0$, $\delta^{LL}_{12}=0.0001$. Besides, parameters $m_{L}$, $m_{E}$, $m_{LL}$ and $m_{EE}$ undergo significant variations. It can be seen that the parameters ($m_{L}$, $m_{E}$) located in the diagonal elements of the slepton(sneutrino) mass matrix depress $Br(\mu \rightarrow e \gamma)$, and the off-diagonal elements($m_{LL}$, $m_{EE}$) produce positive effects on $Br(\mu \rightarrow e \gamma)$. The parameters are constrained by the present limit:\begin{eqnarray}
&&m_{L}>1.05\mathrm{TeV}, m_{E}>0.7\mathrm{TeV}, m_{LL}<1.35\mathrm{TeV}, m_{EE}<2.3\mathrm{TeV}.
\end{eqnarray}
Above all, we find that $\mu \rightarrow e \gamma$ has strict experimental constraints, which cause the parameters $\delta^{AB}_{ij}(A, B=L, R)$, $m_{L}$, $m_{E}$, $m_{LL}$ and $m_{EE}$ to be strictly affected. On this basis, we restudy the  processes $Z\rightarrow l^{\pm}_{i}l^{\mp}_{j}$.
\subsection{$Z\rightarrow l^{\pm}_{i}l^{\mp}_{j}$}
The latest upper limits on the LFV branching ratio of $Z\rightarrow e \mu$, $Z\rightarrow e \tau$ and $Z\rightarrow \mu \tau$ at $95\%$ C.L. are \cite{ATLAS:2022uhq,ATLAS:2021bdj}:
 \begin{eqnarray}
   Br(Z \rightarrow e \mu)<2.62 \times 10^{-7},~~Br(Z \rightarrow e \tau)<5.0 \times 10^{-6},~~Br(Z \rightarrow \mu \tau)<6.5 \times 10^{-6}.
 \end{eqnarray}
 Under the constraints of $(g-2)_{\mu}$ and $l^{-}_{j} \rightarrow l^{-}_{i} \gamma$, we research the branching ratios of the LFV processes $Z\rightarrow l^{\pm}_{i}l^{\mp}_{j}$.
\subsubsection{$Z\rightarrow e \mu$}
\begin{figure}[t]
  \centering
  \includegraphics[width=5.4cm]{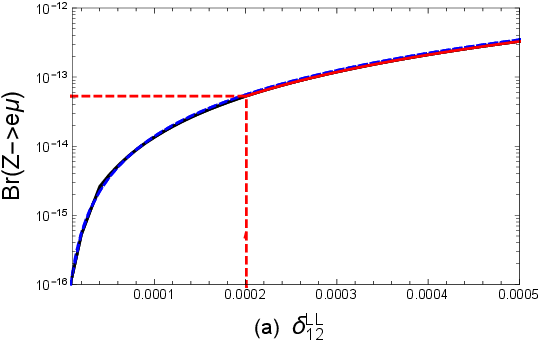}
  \includegraphics[width=5.4cm]{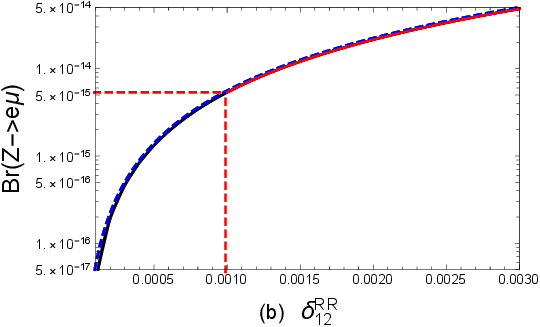}
  \includegraphics[width=5.4cm]{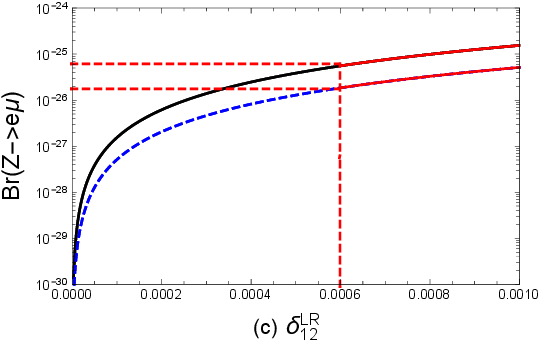}
  \caption{The LFV rates for $Z\rightarrow e \mu$ versus $\delta^{LL}_{12}$, $\delta^{RR}_{12}$, $\delta^{LR}_{12}$, where the solid black line represents the results in the mass eigenstate basis, the dashed blue line represents the results in the electroweak interaction basis. The red solid line is consistent with the present limit of $l^{-}_{j} \rightarrow l^{-}_{i} \gamma$.}
\label{AB12}
\end{figure}
\begin{figure}[t]
  \centering
  \includegraphics[width=6.6cm]{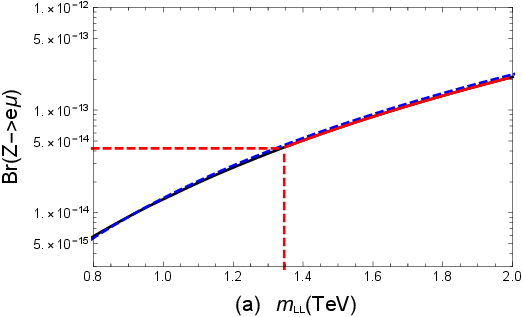}
  \includegraphics[width=6.6cm]{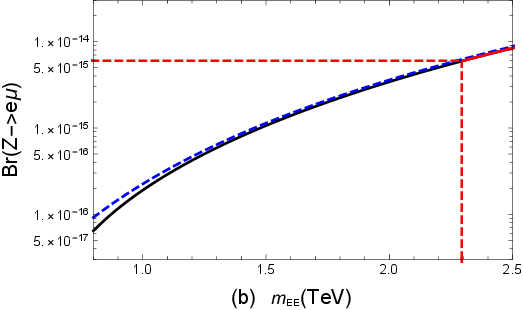}
  \caption{The LFV rates for $Z\rightarrow e \mu$ versus $m_{LL}$ and $m_{EE}$, where the solid black line represents the results in the mass eigenstate basis, the dashed blue line represents the results in the electroweak interaction basis. The red solid line is consistent with the present limit of $l^{-}_{j} \rightarrow l^{-}_{i} \gamma$.}
  \label{figmll}
\end{figure}
In general, the non-diagonal elements of the slepton (sneutrino) mass matrix are considered to be the origin of flavor violation\cite{Calibbi:2017uvl}. In Fig.\ref{AB12}, we plot the contributions from the LFV process $Z\rightarrow e \mu$ changing with the parameters $\delta^{LL}_{12}$, $\delta^{RR}_{12}$ and $\delta^{LR}_{12}$ individually. By comparing the calculation results of the two methods, it can be concluded that the results from parameters $\delta^{LL}_{12}$ and $\delta^{RR}_{12}$ are similarly the same, while the difference from parameter $\delta^{LR}_{12}$ in two methods is about one order. However, due to the small contribution from parameter $\delta^{LR}_{12}$, this difference can be ignored. Comparing these three parameters, we find that the effect from $\delta^{LR}_{12}$ is much smaller than that of $\delta^{RR}_{12}$ and $\delta^{LL}_{12}$, which can confirm the result obtained by approximating analysis in Section III. We can see that the main contributions calculated in two different eigenstates are very similar, and the accuracy of the MIA results is verified. Besides, it can be observed that the branching ratios increase obviously as parameters $\delta^{AB}_{12}(A, B=L, R)$ increase. Therefore, we deduce that parameters $\delta^{AB}_{12}(A, B=L, R)$ are sensitive parameters and have a strong effect on the LFV. The branching ratio $Br(Z\rightarrow e \mu)$ is lower than $10^{-13}$ due to the constraints of $l^{-}_{j} \rightarrow l^{-}_{i} \gamma$.

The parameters $m_{LL}$ and $m_{EE}$ located in the non-diagonal elements of slepton (sneutrino) matrices also have a large effect on the LFV. We show $Br(Z\rightarrow e \mu)$ versus $m_{LL}$ and $m_{EE}$ in the Fig.\ref{figmll}. In Fig.\ref{figmll}(a), we set $\delta^{RR}_{12}=\delta^{LR}_{12}=0$, $\delta^{LL}_{12}=0.0001$. In Fig.\ref{figmll}(b), we set $\delta^{RR}_{12}=0.0002$, $\delta^{LL}_{12}=\delta^{LR}_{12}=0$. It can be observed that both parameters show clear trends and have a positive impact on the process $Z\rightarrow e \mu$. This indicates the parameters $m_{LL}$ and $m_{EE}$ enhance the LFV effect. Under the constraints of $l^{-}_{j} \rightarrow l^{-}_{i} \gamma$, the branching ratio can reach the order around $10^{-14}$.
\begin{figure}[t]
  \centering
  \includegraphics[width=6.6cm]{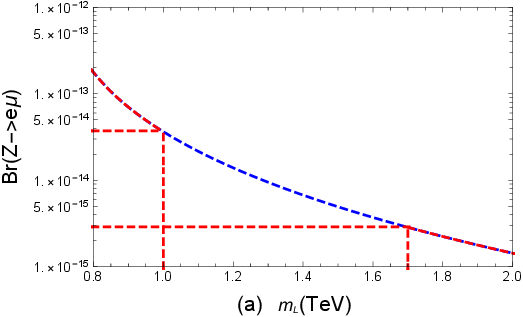}
  \includegraphics[width=6.6cm]{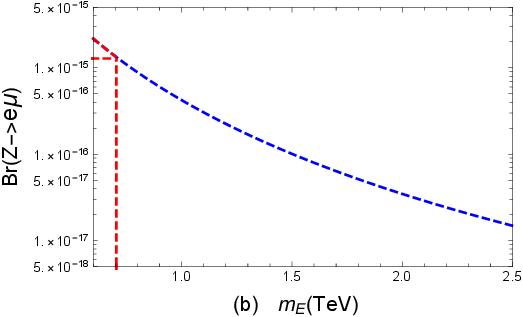}
  \caption{The LFV rates for $Z\rightarrow e \mu$ versus $m_{L}$ and $m_{E}$, where the dashed blue line represents the results in the electroweak interaction basis. The dashed red line indicates exclusion of $(g-2)_{\mu}$ or $l^{-}_{j} \rightarrow l^{-}_{i} \gamma$.}
  \label{ML12}
\end{figure}

The parameters $m_{L}$ and $m_{E}$, which exist in all one-loop functions in the electroweak interaction basis, are the diagonal elements of soft breaking slepton (sneutrino) mass matrices $m^{2}_{\tilde{L},\tilde{R}}$ and associated with the slepton (sneutrino) masses in the MIA. We set $\delta^{RR}_{12}=\delta^{LR}_{12}=0$, $\delta^{LL}_{12}=0.0001$ in Fig.\ref{ML12}(a) and $\delta^{RR}_{12}=0.0002$, $\delta^{LL}_{12}=\delta^{LR}_{12}=0$ in Fig.\ref{ML12}(b), and study the branching ratios of $Z\rightarrow e \mu$ changing with $m_{L}$ and $m_{E}$, respectively. The numerical results indicate that when parameter $m_{L}$ or $m_{E}$ increase, the branching ratio decreases. This indicates the diagonal elements of the slepton (sneutrinos) matrices suppress the LFV effect. The numerical results also show that the contribution of $Br(Z\rightarrow e \mu)$ changing with parameter $m_{L}$ is limited to $10^{-15} \sim 10^{-14}$ under the constraints of $(g-2)_{\mu}$ or $l^{-}_{j} \rightarrow l^{-}_{i} \gamma$, while the contribution of parameter $m_{E}$ is less than $10^{-15}$ under the constraints of $l^{-}_{j} \rightarrow l^{-}_{i} \gamma$.
\subsubsection{$Z\rightarrow e \tau$ and $Z\rightarrow \mu \tau$}
\begin{figure}[t]
  \centering
  \includegraphics[width=5.7cm]{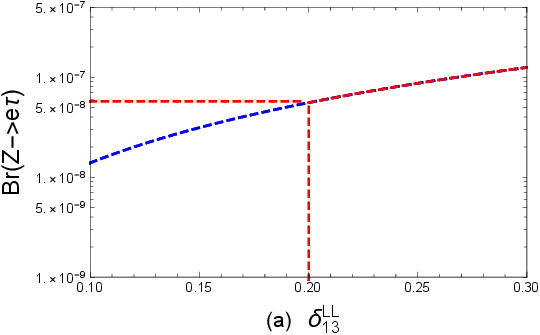}
  \includegraphics[width=5.7cm]{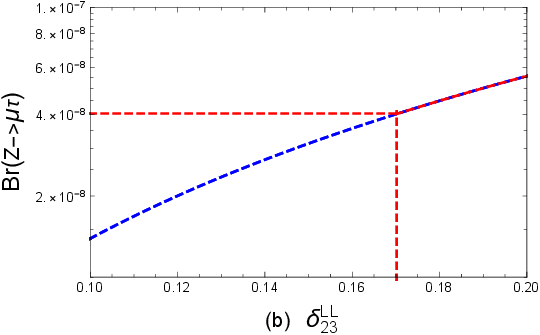}
  \caption{The LFV rates for $Z\rightarrow e \tau$ and $Z\rightarrow \mu \tau$ versus $\delta^{LL}_{13}$, $\delta^{LL}_{23}$. The dashed red line indicates exclusion by $l^{-}_{j} \rightarrow l^{-}_{i} \gamma$.}
  \label{LL13}
\end{figure}
\begin{figure}[t]
  \centering
  \includegraphics[width=5.7cm]{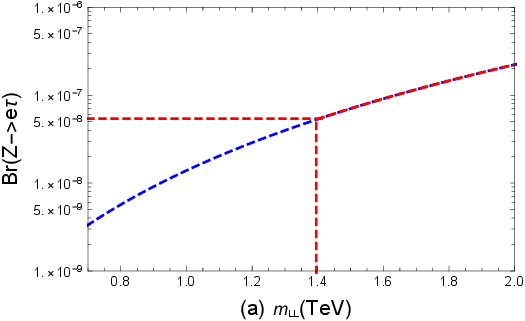}
  \includegraphics[width=5.7cm]{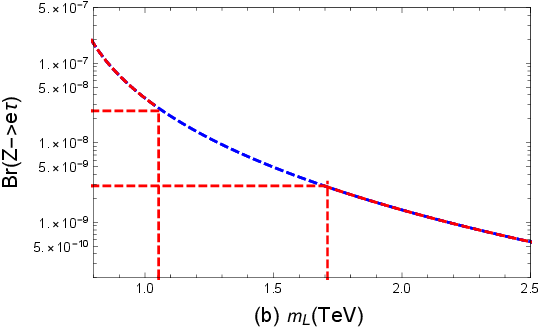}
  \caption{Taking $\delta^{LL}_{13}=0.1$, $\delta^{LR}_{13}=\delta^{RR}_{13}=0$, we plot the LFV rates for $Z\rightarrow e \tau$ versus $m_{LL}$ and $m_{L}$, where the dashed blue line represents the results in the electroweak interaction basis. The dashed red line indicates exclusion of $\tau \rightarrow e \gamma$.  }
  \label{MLL13}
\end{figure}
\begin{figure}[t]
  \centering
  \includegraphics[width=5.7cm]{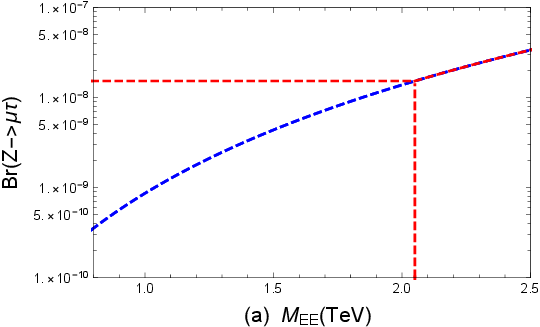}
  \includegraphics[width=5.7cm]{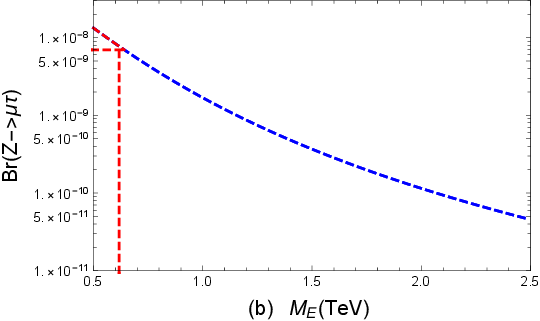}
  \caption{Taking $\delta^{RR}_{23}=0.3$, $\delta^{LR}_{23}=\delta^{LL}_{23}=0$, we show the LFV rates for $Z\rightarrow \mu \tau$ versus $m_{EE}$ and $m_{E}$, where the dashed blue line represents the results in the electroweak interaction basis. The dashed red line indicates exclusion of $\tau \rightarrow \mu \gamma$.}
  \label{MEE23}
\end{figure}

In this subsection, we study the LFV processes $Z\rightarrow e \tau$ and $Z\rightarrow \mu \tau$, as detailed in FIG.\ref{LL13}, FIG.\ref{MLL13} and FIG.\ref{MEE23}. 
 The numerical results show that the diagonal elements($m_{E}$, $m_{L}$) of the slepton(sneutrino) mass matrix increase the LFV effect, while off-diagonal elements($\delta^{LL}_{13}$, $\delta^{LL}_{23}$, $m_{EE}$, $m_{LL}$) decrease the LFV effect. 
In FIG.\ref{LL13}, we plot the $Br(Z\rightarrow e \tau)$ and $Br(Z\rightarrow \mu \tau)$ versus parameters $\delta^{LL}_{13}$ and $\delta^{LL}_{23}$. Under the present limits of $l^{-}_{j} \rightarrow l^{-}_{i} \gamma$,  the parameters $\delta^{LL}_{13}$ and $\delta^{LL}_{23}$ are respectively limited to: $\delta^{LL}_{13}<0.2$, $\delta^{LL}_{23}<0.17$, and $Br(Z\rightarrow e \tau)$ and $Br(Z\rightarrow \mu \tau)$ can be both approximate to $10^{-8}$. As $\delta^{LL}_{13}=0.1, \delta^{RR}_{23}=0.3$, we study the $Br(Z\rightarrow e \tau)$ and $Br(Z\rightarrow \mu \tau)$ changing with paramenters $m_{LL}$, $m_{L}$, $m_{EE}$ and $m_{E}$ in Fig.\ref{MLL13} and Fig.\ref{MEE23}. Under the present limits of $l^{-}_{j} \rightarrow l^{-}_{i} \gamma$ and $(g-2)_{\mu}$, the branching ratios of $Z\rightarrow e \tau$ and $Z\rightarrow \mu \tau$ can be in the order around $5\times10^{-8}$. In particular, the value of parameter $m_{L}$ is constrained to be between $1\mathrm{TeV}$ and $1.7\mathrm{TeV}$, and the $Br(Z\rightarrow e \tau)$ is corresponding in the range of $3\times10^{-9}\sim5\times10^{-8}$.

\section{discussion and conclusion}
In this paper, we study the LFV process $Z\rightarrow l^{\pm}_{i}l^{\mp}_{j}$ in the B-LSSM. We calculate this process in the mass eigenstate basis and the electroweak interaction basis, respectively. Comparing the results obtained by the two methods, the two curves have the same change trends and the corresponding main contributions are almost in the same order. The advantageous aspect of the MIA lies in its ability to reveal the functional correlation between the parameter and the associated branching ratio at the analytical level. When we analyze a certain parameter, the function of the branching ratio changing with the parameter can be precisely found out easily by their analytical equations.

We find that the parameters $\delta^{AB}_{ij}(A,B=L,R)$, $m_{L}$, $m_{E}$, $m_{EE}$ and $m_{LL}$ in the slepton (sneutrino) mass matrix have a significant effect on the $Br(Z\rightarrow l^{\pm}_{i}l^{\mp}_{j})$. Here, $\delta^{AB}_{12}(A,B=L,R)$, $m_{LL}$ and $m_{EE}$ present in the off-diagonal of the slepton (sneutrinos) mass matrix, whose increase enlarge the branching ratios of $Z\rightarrow l^{\pm}_{i}l^{\mp}_{j}$. As the diagonal terms of slepton (sneutrino) mass matrix, the increase of $m_{L}$ and $m_{E}$ decrease the branching ratios of $Z\rightarrow l^{\pm}_{i}l^{\mp}_{j}$. Therefore, the diagonal elements of the slepton (sneutrino) mass matrix suppress the LFV effect, while the off-diagonal elements enhance the LFV effect. This property is similar to other LFV processes(e.g., $l^{-}_{j} \rightarrow l^{-}_{i} \gamma$).

Through the numerical analysis, we find that $(g-2)_{\mu}$ and $l^{-}_{j} \rightarrow l^{-}_{i} \gamma$ have certain constraints on the parameter spaces, and the parameters in the slepton (sneutrinos) mass matrix are the most stringent, such as parameters $\delta^{AB}_{ij}(A, B= L, R)$, $m_{EE}$, $m_{E}$, $m_{LL}$ and $m_{L}$ are constrained. Considering the constraints of $(g-2)_{\mu}$ and $l^{-}_{j} \rightarrow l^{-}_{i} \gamma$, the $Br(Z\rightarrow e \mu)$ can be in the order around $10^{-14}$, and the upper bounds of $Br(Z\rightarrow e \tau)$ and $Br(Z\rightarrow \mu \tau)$ can reach $5\times 10^{-8}$. More specifically, as the parameter $m_{L}$ is restricted between $1\mathrm{TeV}$ and $1.7\mathrm{TeV}$,  the branching ratios of $Z\rightarrow l^{\pm}_{i}l^{\mp}_{j}$ that are strictly limited as: $3\times10^{-15}<Br(Z\rightarrow e \mu)<4\times10^{-14}$ and $3\times10^{-9}<Br(Z\rightarrow e \tau)<5\times10^{-8}$.  The numercial results discussed above all satisfy the present constraints of $Z\rightarrow l^{\pm}_{i}l^{\mp}_{j}$. We hope that the experimental results for $Z\rightarrow l^{\pm}_{i}l^{\mp}_{j}$ can be detected in the future, which may provide the possibilities for finding NP beyond the SM.

{\bf Acknowledgments}
This work is supported by the Major Project of National Natural Science Foundation of China (NNSFC) (No. 12235008), the National Natural Science Foundation of China (NNSFC) (No. 12075074, No. 12075073), the Natural Science Foundation of Hebei province(No.A2022201022, No. A2020201002, No. A2023201041, No. A2022201017), the Natural Science Foundation of Hebei Education Department(No. QN2022173).

\appendix
\section{The one-loop functions} \label{OLF}
 The one-loop functions $I_{2}(x, y, z)$, $G_{1}(x, y, z)$, $G_{2}(x, y, z)$, $G_{3}(x, y, z)$, $G_{4}(x, y, z, t)$, $G_{5}(x, y, z, t)$, $G_{6}(x, y, z, t, n)$ and $G_{7}(x, y, z, t, n)$ can be written as
\begin{eqnarray*}
&&I_{2}(x, y, z)=\frac{1}{16 \pi^{2}}\Big[\frac{1+\ln x}{(z-x)(y-x)}-\frac{x\ln x}{(x-y)^{2}(x-z)}-\frac{x\ln x}{(x-z)^{2}(x-y)} \nonumber\\
&&~~~~~~~~~~~~~~~~~-\frac{y\ln y}{(x-y)^{2}(z-y)}-\frac{z\ln z}{(x-y)^{2}(y-z)}\Big], \\
&&G_{1}(x, y, z)=\frac{1}{16 \pi^{2}}\Big[\frac{x\ln x}{(z-x)(y-x)}+\frac{y\ln y}{(x-y)(z-y)}+\frac{z\ln z}{(x-z)(y-z)}\Big], \\
&&G_{2}(x, y, z)=\frac{1}{16 \pi^{2}}\Big[\frac{x^{2}\ln x}{(z-x)(y-x)}+\frac{y^{2}\ln y}{(x-y)(z-y)}+\frac{z^{2}\ln z}{(x-z)(y-z)}\Big], \\
&&G_{3}(x, y, z)=\frac{1}{16 \pi^{2}}\Big[\frac{x+2x\ln x}{(z-x)(y-x)}+\frac{x^{2}\ln x}{(z-x)(y-x)^{2}}+\frac{x^{2}\ln x}{(y-x)(z-x)^{2}} \nonumber\\
&&~~~~~~~~~~~~~~~~~+\frac{y^{2}\ln y}{(y-z)(x-y)^{2}}+\frac{z^{2}\ln z}{(y-z)(x-z)^{2}}\Big], \\
&&G_{4}(x, y, z, t)=\frac{1}{16 \pi^{2}}\Big[\frac{2x\ln x}{(x-y)(x-z)}-\frac{2x\ln x}{(x-t)(x-y)}-\frac{x^{2}\ln x}{(x-z)(x-y)^{2}}+\frac{x^{2}\ln x}{(x-t)(x-y)^{2}} \nonumber\\
&&~~~~~~~~~~~~~~~~-\frac{x^{2}\ln x}{(x-y)(x-z)^{2}}+\frac{x^{2}\ln x}{(x-y)(x-t)^{2}}-\frac{y^{2}\ln y}{(z-y)(x-y)^{2}}+\frac{y^{2}\ln y}{(t-y)(x-y)^{2}} \nonumber\\
&&~~~~~~~~~~~~~~~~-\frac{z^{2}\ln z}{(y-z)(x-z)^{2}}+\frac{t^{2}\ln t}{(y-t)(x-t)^{2}}\Big]\frac{1}{z-t}, \\
&&G_{5}(x, y, z, t)=\frac{1}{16 \pi^{2}}\Big[\frac{1+\ln x}{(x-y)(x-z)}-\frac{1+\ln x}{(x-t)(x-y)}-\frac{x\ln x}{(x-z)(x-y)^{2}}+\frac{x\ln x}{(x-t)(x-y)^{2}} \nonumber\\
&&~~~~~~~~~~~~~~~~-\frac{x\ln x}{(x-y)(x-z)^{2}}+\frac{x\ln x}{(x-y)(x-t)^{2}}-\frac{y\ln y}{(z-y)(x-y)^{2}}+\frac{y\ln y}{(t-y)(x-y)^{2}} \nonumber\\
&&~~~~~~~~~~~~~~~~-\frac{z\ln z}{(y-z)(x-z)^{2}}+\frac{t\ln t}{(y-t)(x-t)^{2}}\Big]\frac{1}{z-t},  \\
&&G_{6}(x, y, z, t, n)=[G_{4}(x, y, z, t)-G_{4}(x, y, z, n)]\frac{1}{t-n}, \\
&&G_{7}(x, y, z, t, n)=[G_{5}(x, y, z, t)-G_{5}(x, y, z, n)]\frac{1}{t-n}. 
\end{eqnarray*}

\section{The couplings in the mass eigenstate basis}
\begin{eqnarray*}
&&H^{Z\chi^{0}_{i}\chi^{0}_{j}}_{L}=-\frac{i}{2}(N^{\ast}_{j,3}(g_{1}\sin \theta_{W} \cos \theta_{W}^{\prime}+g_{2}\cos \theta_{W} \cos \theta_{W}^{\prime}-g_{YB}\sin \theta_{W}^{\prime})N_{i,3} \nonumber\\
&&~~~~~~~~~~~~~-N^{\ast}_{j,4}(g_{1}\sin \theta_{W} \cos \theta_{W}^{\prime}+g_{2}\cos \theta_{W} \cos \theta_{W}^{\prime}-g_{YB}\sin \theta_{W}^{\prime})N_{i,4} \nonumber\\
&&~~~~~~~~~~~~~-2g_{B}\sin \theta_{W}^{\prime}(N^{\ast}_{j,6}N_{i,6}-N^{\ast}_{j,7}N_{i,7})) \\
&&H^{Z\chi^{0}_{i}\chi^{0}_{j}}_{R}=\frac{i}{2}(N^{\ast}_{i,3}(g_{1}\sin \theta_{W} \cos \theta_{W}^{\prime}+g_{2}\cos \theta_{W} \cos \theta_{W}^{\prime}-g_{YB}\sin \theta_{W}^{\prime})N_{j,3} \nonumber\\
&&~~~~~~~~~~~~~-N^{\ast}_{i,4}(g_{1}\sin \theta_{W} \cos \theta_{W}^{\prime}+g_{2}\cos \theta_{W} \cos \theta_{W}^{\prime}-g_{YB}\sin \theta_{W}^{\prime})N_{j,4} \nonumber\\
&&~~~~~~~~~~~~~-2g_{B}\sin \theta_{W}^{\prime}(N^{\ast}_{i,6}N_{j,6}-N^{\ast}_{i,7}N_{j,7})) \\
&&H^{Z\chi^{\pm}_{i}\chi^{\pm}_{j}}_{L}=\frac{i}{2}(2g_{2}U^{\ast}_{j,1}\cos \theta \cos \theta^{\prime}U_{i,1} \nonumber\\
&&~~~~~~~~~~~~~+U^{\ast}_{j,2}(-g_{1}\sin \theta \cos \theta^{\prime}+g_{2}\cos \theta \cos \theta^{\prime}+g_{YB}\sin \theta^{\prime})U_{i,2}) \\
&&H^{Z\chi^{\pm}_{i}\chi^{\pm}_{j}}_{R}=\frac{i}{2}(2g_{2}V^{\ast}_{i,1}\cos \theta \cos \theta^{\prime}V_{j,1} \nonumber\\
&&~~~~~~~~~~~~~+V^{\ast}_{i,2}(-g_{1}\sin \theta \cos \theta^{\prime}+g_{2}\cos \theta \cos \theta^{\prime}+g_{YB}\sin \theta^{\prime})V_{j,2}) \\
&&H^{Z\tilde{\nu}^{I}_{i}\tilde{\nu}^{R}_{j}}=\frac{1}{2}(g_{1}\cos \theta^{\prime} \sin \theta +g_{2}\cos \theta \cos \theta^{\prime}-(g_{YB}+g_{B})\sin \theta^{\prime})\sum^{3}_{j_{1}=1}Z^{I,\ast}_{i,j_{1}}Z^{R,\ast}_{j,j_{1}} \nonumber\\
&&~~~~~~~~~~~~~-g_{B}\sin \theta^{\prime}\sum^{3}_{j_{1}=1}Z^{I,\ast}_{i3+j_{1}}Z^{R,\ast}_{j3+j_{1}} \\
&&H^{Z\tilde{l}_{i}\tilde{l}_{j}}=\frac{i}{2}(-g_{1}\sin \theta \cos \theta^{\prime}+g_{2}\cos \theta \cos \theta^{\prime}+(g_{YB}+g_{B})\sin \theta^{\prime})\sum^{3}_{j_{1}=1}Z^{E,\ast}_{i,j_{1}}Z^{E}_{j,j_{1}} \nonumber\\
&&~~~~~~~~~~~-(2g_{1}\sin \theta \cos \theta^{\prime}-(2g_{YB}+g_{B})\sin \theta^{\prime})\sum^{3}_{j_{1}=1}Z^{E,\ast}_{i,3+j_{1}}Z^{E}_{j,3+j_{1}} \\
&&H^{l_{j}\tilde{l}_{k}\tilde{\chi}^{0}_{i}}_{L}=\frac{i}{2}(\sqrt{2}g_{1}N^{\ast}_{i,1}Z^{E}_{k,j}+\sqrt{2}g_{2}N^{\ast}_{i,2}Z^{E}_{k,j}+\sqrt{2}g_{YB}N^{\ast}_{i,5}Z^{E}_{k,j} \nonumber\\
&&~~~~~~~~~~~~+\sqrt{2}g_{B}N^{\ast}_{i,5}Z^{E}_{k,j}+2N^{\ast}_{i,3}Y_{e,jj}Z^{E}_{k,3+j}) \\
&&H^{l_{j}\tilde{l}_{k}\tilde{\chi}^{0}_{i}}_{R}=i(-\frac{1}{\sqrt{2}}Z^{E}_{k,3+j}(2g_{1}N_{i1}+(2g_{YB}+g_{B})N_{i,5})-Y^{\ast}_{e,jj}Z^{E}_{k,j}N_{i,3}) \\
&&H^{l_{j}\tilde{\nu}^{I}_{k}\chi^{\pm}_{i}}_{L}=\frac{1}{\sqrt{2}}(-g_{2}V^{\ast}_{i,1}Z^{I,\ast}_{k,j}+V^{\ast}_{i,2}Z^{I,\ast}_{k,3+j}Y^{\ast}_{\nu,jj}) \\
&&H^{l_{j}\tilde{\nu}^{I}_{k}\chi^{\pm}_{i}}_{R}=\frac{1}{\sqrt{2}}(Z^{I,\ast}_{k,j}Y^{\ast}_{e,jj}U_{i,2}) \\
&&H^{l_{j}\tilde{\nu}^{R}_{k}\chi^{\pm}_{i}}_{L}=i\frac{1}{\sqrt{2}}(-g_{2}V^{\ast}_{i,1}Z^{R,\ast}_{k,j}+V^{\ast}_{i,2}Z^{R,\ast}_{k,3+j}Y^{\ast}_{\nu,jj}) \\
&&H^{l_{j}\tilde{\nu}^{R}_{k}\chi^{\pm}_{i}}_{R}=i\frac{1}{\sqrt{2}}(Z^{R,\ast}_{kj}Y^{\ast}_{e,jj}U_{i,2}) \\
\end{eqnarray*}
\begin{eqnarray*}
H^{\bar{l}_{j}\tilde{l}_{k}\tilde{\chi}^{0}_{i}}_{R}=[H^{l_{j}\tilde{l}_{k}\tilde{\chi}^{0}_{i}}_{L}]^{*},  H^{\bar{l}_{j}\tilde{l}_{k}\tilde{\chi}^{0}_{i}}_{L}=[H^{l_{j}\tilde{l}_{k}\tilde{\chi}^{0}_{i}}_{R}]^{*}, H^{\bar{l}_{j}\tilde{\nu}^{I}_{k}\chi^{\pm}_{i}}_{R}=[H^{l_{j}\tilde{\nu}^{I}_{k}\chi^{\pm}_{i}}_{L}]^{*}, \nonumber\\ H^{\bar{l}_{j}\tilde{\nu}^{I}_{k}\chi^{\pm}_{i}}_{L}=[H^{l_{j}\tilde{\nu}^{I}_{k}\chi^{\pm}_{i}}_{R}]^{*},
H^{\bar{l}_{j}\tilde{\nu}^{R}_{k}\chi^{\pm}_{i}}_{R}=[H^{l_{j}\tilde{\nu}^{R}_{k}\chi^{\pm}_{i}}_{L}]^{*},  H^{\bar{l}_{j}\tilde{\nu}^{R}_{k}\chi^{\pm}_{i}}_{L}=[H^{l_{j}\tilde{\nu}^{R}_{k}\chi^{\pm}_{i}}_{R}]^{*}.
\end{eqnarray*}

\section{The relevant Feynman rules in the electroweak interaction basis}
\begin{figure}
\includegraphics[width=14cm]{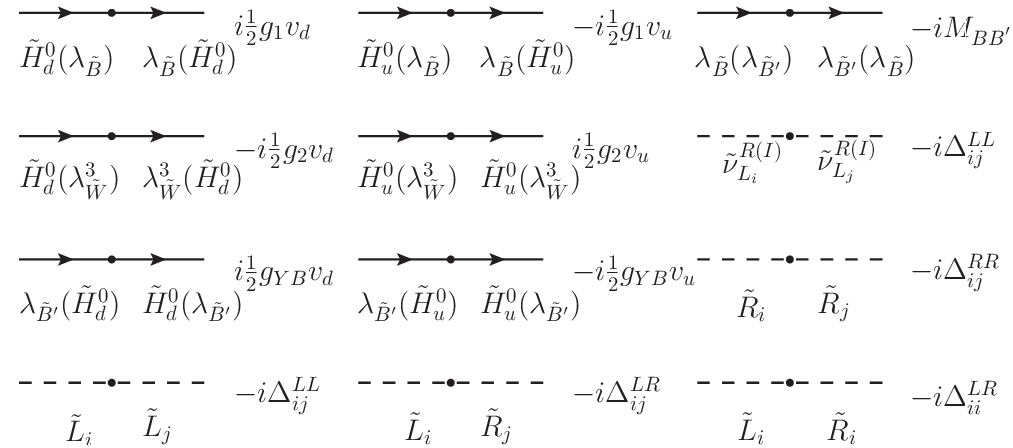}
\includegraphics[width=14cm]{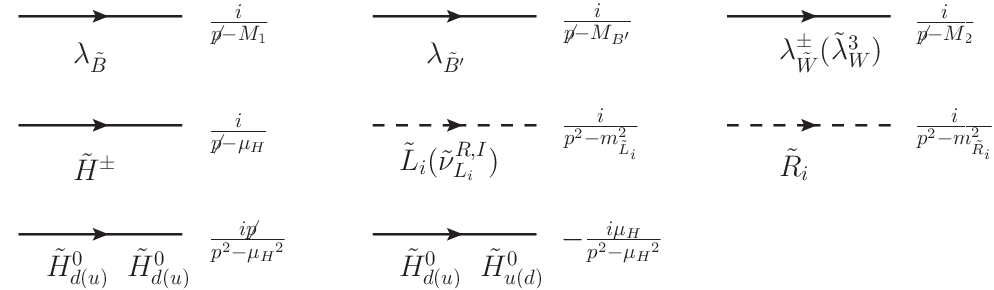}
\caption{Feyman rules for the relevant insertions and  relevant propagators}
\label{1234}
\end{figure}

The relevant Feynman rules for the present computation are collected in Fig.\ref{1234}.

\section{The contributions from Feynman diagrams with the MIA method}\label{MR}
To save space in the text, we use some symbols to represent the coupling vertices associated with the Z boson, as follows:
\begin{eqnarray}
&&C_{L}=-g_{1}\sin \theta \cos \theta^{\prime}+g_{2}\cos \theta \cos \theta^{\prime}+(g_{YB}+g_{B})\sin \theta^{\prime},\\
&&C_{R}=2g_{1}\sin \theta \cos \theta^{\prime}-(2g_{YB}+g_{B})\sin \theta^{\prime},\\
&&E=-\frac{1}{2}(g_{2}\cos \theta \cos \theta^{\prime}-g_{1}\sin \theta \cos \theta^{\prime}+g_{YB}\sin \theta^{\prime}).
\end{eqnarray}

The one-loop contributions from Fig.\ref{Mia1}(1a)
\begin{eqnarray}
&&A^{S_{1}}_{R}(1a)=\frac{1}{2\Lambda ^{3}}m_{l_{i}}g^{2}_{1}\Delta_{ij}^{LR}E[G_{4}(x_{\mu_{H}},x_{\tilde{R}_{j}},x_{\tilde{L}_{i}},x_{1})(\sqrt{x_{1}}-\sqrt{x_{\mu_{H}}}\tan \beta)\nonumber\\
&&~~~~~~~~-2\sqrt{x_{1}}x_{\mu_{H}}G_{5}(x_{\mu_{H}},\tilde{x}_{R_{j}},x_{\tilde{L}_{i}},x_{1})],\\
&&A^{S_{2}}_{R}(1a)=\frac{1}{4\Lambda ^{3}}m_{l_{i}}g_{YB}(g_{B}+2g_{YB})\Delta_{ij}^{LR}E[G_{4}(x_{\mu_{H}},x_{\tilde{R}_{j}},x_{\tilde{L}_{i}},x_{B^{\prime}})(\sqrt{x_{B^{\prime}}}-\sqrt{x_{\mu_{H}}}\tan \beta)\nonumber\\
&&~~~~~~~~-2\sqrt{x_{B^{\prime}}}x_{\mu_{H}}G_{5}(x_{\mu_{H}},x_{\tilde{R}_{j}},x_{\tilde{L}_{i}},x_{\tilde{B}^{\prime}})].
\end{eqnarray}

The one-loop contributions from Fig.\ref{Mia1}(1b)
\begin{eqnarray}
&&A^{S_{1}}_{L}(1b)=\frac{1}{4\Lambda ^{3}}m_{l_{i}}g^{2}_{1}\Delta_{ij}^{LR}E[G_{4}(x_{\mu_{H}},x_{\tilde{L}_{j}},x_{\tilde{R}_{i}},x_{1})(\sqrt{x_{1}}-\sqrt{x_{\mu_{H}}}\tan \beta)\nonumber\\
&&~~~~~~~-2\sqrt{x_{1}}x_{\mu_{H}}G_{5}(x_{\mu_{H}},x_{\tilde{L}_{j}},x_{\tilde{R}_{i}},x_{1})],\\
&&A^{S_{2}}_{L}(1b)=\frac{1}{4\Lambda ^{3}}m_{l_{i}}g_{YB}(g_{B}+g_{YB})\Delta_{ij}^{LR}E[G_{4}(x_{\mu_{H}},x_{\tilde{L}_{j}},x_{\tilde{R}_{i}},x_{B^{\prime}})(\sqrt{x_{B^{\prime}}}-\sqrt{x_{\mu_{H}}}\tan \beta)\nonumber\\
&&~~~~~~~-2\sqrt{x_{B^{\prime}}}x_{\mu_{H}}G_{5}(x_{\mu_{H}},x_{\tilde{L}_{j}},x_{\tilde{R}_{i}},x_{B^{\prime}})],\\
&&A^{S_{3}}_{L}(1b)=-\frac{1}{4{\Lambda ^{3}}}m_{l_{i}}g^{2}_{2}\Delta_{ij}^{LR}E[G_{4}(x_{\mu_{H}},x_{\tilde{L}_{j}},x_{R_{i}},x_{2})(\sqrt{x_{2}}-\sqrt{x_{\mu_{H}}}\tan \beta)\nonumber\\
&&~~~~~~~-2\sqrt{x_{2}}x_{\mu_{H}}G_{5}(x_{\mu_{H}},x_{\tilde{L}_{j}},x_{\tilde{R}_{i}},x_{2})],
\end{eqnarray}

The one-loop contributions from Fig.\ref{Mia1}(1c)
\begin{eqnarray}
&&A^{S_{1}}_{R}(1c)=\frac{1}{2{\Lambda ^{3}}}m_{l_{j}}g^{2}_{1}\Delta_{ij}^{LR}E[G_{4}(x_{\mu_{H}},x_{\tilde{L}_{j}},x_{\tilde{R}_{i}},x_{1})(\sqrt{x_{1}}-\sqrt{x_{\mu_{H}}}\tan \beta)\nonumber\\
&&~~~~~~~-2\sqrt{x_{1}}x_{\mu_{H}}G_{5}(x_{\mu_{H}},x_{\tilde{L}_{j}},x_{\tilde{R}_{i}},x_{1})],\\
&&A^{S_{2}}_{R}(1c)=\frac{1}{4{\Lambda ^{3}}}m_{l_{j}}g_{YB}(g_{B}+2g_{YB})\Delta_{ij}^{LR}E[G_{4}(x_{\mu_{H}},x_{\tilde{L}_{j}},x_{\tilde{R}_{i}},x_{B^{'}})(\sqrt{x_{B^{\prime}}}-\sqrt{x_{\mu_{H}}}\tan \beta)\nonumber\\
&&~~~~~~~-2\sqrt{x_{B^{\prime}}}x_{\mu_{H}}G_{5}(x_{\mu_{H}},x_{\tilde{L}_{j}},x_{\tilde{R}_{i}},x_{B^{\prime}})].
\end{eqnarray}

The one-loop contributions from Fig.\ref{Mia1}(1d)
\begin{eqnarray}
&&A^{S_{1}}_{L}(1d)=\frac{1}{4\Lambda ^{3}}m_{l_{j}}g^{2}_{1}\Delta_{ij}^{LR}E[G_{4}(x_{\mu_{H}},x_{\tilde{R}_{j}},x_{L_{i}},x_{1})(\sqrt{x_{1}}-\sqrt{x_{\mu_{H}}}\tan \beta)\nonumber\\
&&~~~~~~~-2\sqrt{x_{1}}x_{\mu_{H}}G_{5}(x_{\mu_{H}},x_{\tilde{R}_{j}},x_{\tilde{L}_{i}},x_{1})],\\
&&A^{S_{2}}_{L}(1d)=\frac{1}{4\Lambda ^{3}}m_{l_{j}}g_{YB}(g_{B}+g_{YB})\Delta_{ij}^{LR}E[G_{4}(x_{\mu_{H}},x_{\tilde{R}_{j}},x_{\tilde{L}_{i}},x_{B^{\prime}})(\sqrt{x_{B^{\prime}}}-\sqrt{x_{\mu_{H}}}\tan \beta)\nonumber\\
&&~~~~~~~-2\sqrt{x_{B^{\prime}}}x_{\mu_{H}}G_{5}(x_{\mu_{H}},x_{\tilde{R}_{j}},x_{\tilde{L}_{i}},x_{B^{\prime}})],\\
&&A^{S_{3}}_{L}(1d)=-\frac{1}{4\Lambda ^{3}}m_{l_{j}}g^{2}_{2}\Delta_{ij}^{LR}E[G_{4}(x_{\mu_{H}},x_{\tilde{R}_{j}},x_{\tilde{L}_{i}},x_{2})(\sqrt{x_{2}}-\sqrt{x_{\mu_{H}}}\tan \beta)\nonumber\\
&&~~~~~~~-2\sqrt{x_{2}}x_{\mu_{H}}G_{5}(x_{\mu_{H}},x_{\tilde{R}_{j}},x_{\tilde{L}_{i}},x_{2})].
\end{eqnarray}

The one-loop contributions from Fig.\ref{Mia1}(1e)
\begin{eqnarray}
&&A^{S_{4}}_{R}(1e)=-\frac{1}{2\Lambda ^{2}}Y_{e,ii}Y_{e,jj}\Delta_{ij}^{LL}E[G_{3}(x_{\mu_{H}},x_{\tilde{L}_{j}},x_{\tilde{L}_{i}})-2I_{2}(x_{\mu_{H}},x_{\tilde{L}_{j}},x_{\tilde{L}_{i}})].
\end{eqnarray}

The one-loop contributions from Fig.\ref{Mia1}(1f)
\begin{eqnarray}
&&A^{S_{4}}_{L}(1f)=-\frac{1}{2\Lambda ^{2}}Y_{e,ii}Y_{e,jj}\Delta_{ij}^{RR}E[G_{3}(x_{\mu_{H}},x_{\tilde{R}_{j}},x_{\tilde{R}_{i}})-2I_{2}(x_{\mu_{H}},x_{\tilde{R}_{j}},x_{\tilde{R}_{i}})].
\end{eqnarray}

The one-loop contributions from Fig.\ref{Mia1}(1g)
\begin{eqnarray}
&&A^{S_{1}}_{R}(1g)=-\frac{1}{2\Lambda ^{5}}m_{l_{i}}g^{2}_{1}[G_{6}(x_{\mu_{H}},x_{\tilde{R}_{j}},x_{\tilde{R}_{i}},x_{\tilde{L}_{i}},x_{1})(\sqrt{x_{1}}-\sqrt{x_{\mu_{H}}}\tan \beta)\nonumber\\
&&~~~~~~~~-2\sqrt{x_{1}}x_{\mu_{H}}G_{7}(x_{\mu_{H}},\tilde{x}_{R_{j}},x_{\tilde{R}_{i}},x_{\tilde{L}_{i}},x_{1})]\Delta_{ii}^{LR}\Delta_{ij}^{RR}E,\\
&&A^{S_{2}}_{R}(1g)=-\frac{1}{4\Lambda ^{5}}m_{l_{i}}g_{YB}(g_{B}+2g_{YB})[G_{6}(x_{\mu_{H}},x_{\tilde{R}_{j}},x_{\tilde{L}_{i}},x_{B^{\prime}})(\sqrt{x_{B^{\prime}}}-\sqrt{x_{\mu_{H}}}\tan \beta)\nonumber\\
&&~~~~~~~~-2\sqrt{x_{B^{\prime}}}x_{\mu_{H}}G_{7}(x_{\mu_{H}},\tilde{x}_{R_{j}},x_{\tilde{R}_{i}},x_{\tilde{L}_{i}},x_{\tilde{B}^{\prime}})]\Delta_{ii}^{LR}\Delta_{ij}^{RR}E.
\end{eqnarray}

The one-loop contributions from Fig.\ref{Mia1}(1h)
\begin{eqnarray}
&&A^{S_{1}}_{L}(1h)=-\frac{1}{4\Lambda ^{5}}m_{l_{i}}g^{2}_{1}[G_{6}(x_{\mu_{H}},x_{\tilde{L}_{j}},x_{\tilde{L}_{i}},x_{\tilde{R}_{i}},x_{1})(\sqrt{x_{1}}-\sqrt{x_{\mu_{H}}}\tan \beta)\nonumber\\
&&~~~~~~~-2\sqrt{x_{1}}x_{\mu_{H}}G_{7}(x_{\mu_{H}},x_{\tilde{L}_{j}},x_{\tilde{L}_{i}},x_{\tilde{R}_{i}},x_{1})]\Delta_{ii}^{LR}\Delta_{ij}^{LL}E,\\
&&A^{S_{2}}_{L}(1h)=-\frac{1}{4\Lambda ^{5}}m_{l_{i}}g_{YB}(g_{B}+g_{YB})[G_{6}(x_{\mu_{H}},x_{\tilde{L}_{j}},x_{\tilde{L}_{i}},x_{\tilde{R}_{i}},x_{B^{\prime}})(\sqrt{x_{B^{\prime}}}-\sqrt{x_{\mu_{H}}}\tan \beta)\nonumber\\
&&~~~~~~~-2\sqrt{x_{B^{\prime}}}x_{\mu_{H}}G_{7}(x_{\mu_{H}},x_{\tilde{L}_{j}},x_{\tilde{L}_{i}},x_{\tilde{R}_{i}},x_{B^{\prime}})]\Delta_{ii}^{LR}\Delta_{ij}^{LL}E,\\
&&A^{S_{3}}_{L}(1h)=\frac{1}{4{\Lambda ^{5}}}m_{l_{i}}g^{2}_{2}[G_{6}(x_{\mu_{H}},x_{\tilde{L}_{j}},x_{\tilde{L}_{i}},x_{R_{i}},x_{2})(\sqrt{x_{2}}-\sqrt{x_{\mu_{H}}}\tan \beta)\nonumber\\
&&~~~~~~~-2\sqrt{x_{2}}x_{\mu_{H}}G_{7}(x_{\mu_{H}},x_{\tilde{L}_{j}},x_{\tilde{L}_{i}},x_{\tilde{R}_{i}},x_{2})]\Delta_{ii}^{LR}\Delta_{ij}^{LL}E.
\end{eqnarray}

The one-loop contributions from Fig.\ref{Mia1}(1i)
\begin{eqnarray}
&&A^{S_{1}}_{R}(1i)=-\frac{1}{2{\Lambda ^{5}}}m_{l_{j}}g^{2}_{1}[G_{6}(x_{\mu_{H}},x_{\tilde{L}_{j}},x_{\tilde{L}_{i}},x_{\tilde{R}_{i}},x_{1})(\sqrt{x_{1}}-\sqrt{x_{\mu_{H}}}\tan \beta)\nonumber\\
&&~~~~~~~-2\sqrt{x_{1}}x_{\mu_{H}}G_{7}(x_{\mu_{H}},x_{\tilde{L}_{j}},x_{\tilde{L}_{i}},x_{\tilde{R}_{i}},x_{1})]\Delta_{ii}^{LR}\Delta_{ij}^{LL}E,\\
&&A^{S_{2}}_{R}(1i)=-\frac{1}{4{\Lambda ^{5}}}m_{l_{j}}g_{YB}(g_{B}+2g_{YB})[G_{6}(x_{\mu_{H}},x_{\tilde{L}_{j}},x_{\tilde{L}_{i}},x_{\tilde{R}_{i}},x_{B^{'}})(\sqrt{x_{B^{\prime}}}-\sqrt{x_{\mu_{H}}}\tan \beta)\nonumber\\
&&~~~~~~~-2\sqrt{x_{B^{\prime}}}x_{\mu_{H}}G_{7}(x_{\mu_{H}},x_{\tilde{L}_{j}},x_{\tilde{L}_{i}},x_{\tilde{R}_{i}},x_{B^{\prime}})]\Delta_{ii}^{LR}\Delta_{ij}^{LL}E.
\end{eqnarray}

The one-loop contributions from Fig.\ref{Mia1}(1j)
\begin{eqnarray}
&&A^{S_{1}}_{L}(1j)=-\frac{1}{4\Lambda ^{5}}m_{l_{j}}g^{2}_{1}[G_{6}(x_{\mu_{H}},x_{\tilde{R}_{j}},x_{\tilde{R}_{i}},x_{L_{i}},x_{1})(\sqrt{x_{1}}-\sqrt{x_{\mu_{H}}}\tan \beta)\nonumber\\
&&~~~~~~~-2\sqrt{x_{1}}x_{\mu_{H}}G_{7}(x_{\mu_{H}},x_{\tilde{R}_{j}},x_{\tilde{R}_{i}},x_{\tilde{L}_{i}},x_{1})]\Delta_{ii}^{LR}\Delta_{ij}^{RR}E,\\
&&A^{S_{2}}_{L}(1j)=-\frac{1}{4\Lambda ^{5}}m_{l_{j}}g_{YB}(g_{B}+g_{YB})[G_{6}(x_{\mu_{H}},x_{\tilde{R}_{j}},x_{\tilde{R}_{i}},x_{\tilde{L}_{i}},x_{B^{\prime}})(\sqrt{x_{B^{\prime}}}-\sqrt{x_{\mu_{H}}}\tan \beta)\nonumber\\
&&~~~~~~~-2\sqrt{x_{B^{\prime}}}x_{\mu_{H}}G_{7}(x_{\mu_{H}},x_{\tilde{R}_{j}},x_{\tilde{R}_{i}},x_{\tilde{L}_{i}},x_{B^{\prime}})]\Delta_{ii}^{LR}\Delta_{ij}^{RR}E,\\
&&A^{S_{3}}_{L}(1j)=\frac{1}{4\Lambda ^{5}}m_{l_{j}}g^{2}_{2}[G_{6}(x_{\mu_{H}},x_{\tilde{R}_{j}},x_{\tilde{R}_{i}},x_{\tilde{L}_{i}},x_{2})(\sqrt{x_{2}}-\sqrt{x_{\mu_{H}}}\tan \beta)\nonumber\\
&&~~~~~~~-2\sqrt{x_{2}}x_{\mu_{H}}G_{7}(x_{\mu_{H}},x_{\tilde{R}_{j}},x_{\tilde{R}_{i}},x_{\tilde{L}_{i}},x_{2})]\Delta_{ii}^{LR}\Delta_{ij}^{RR}E.
\end{eqnarray}

The one-loop contributions from Fig.\ref{Mia1}(1k)
\begin{eqnarray}
&&A^{S_{1}}_{R}(1k)=-\frac{1}{2\Lambda ^{5}}m_{l_{i}}g^{2}_{1}[G_{6}(x_{\mu_{H}},x_{\tilde{R}_{j}},x_{\tilde{L}_{j}},x_{\tilde{L}_{i}},x_{1})(\sqrt{x_{1}}-\sqrt{x_{\mu_{H}}}\tan \beta)\nonumber\\
&&~~~~~~~~-2\sqrt{x_{1}}x_{\mu_{H}}G_{7}(x_{\mu_{H}},\tilde{x}_{R_{j}},x_{\tilde{L}_{j}},x_{\tilde{L}_{i}},x_{1})]\Delta_{jj}^{LR}\Delta_{ij}^{LL}E,\\
&&A^{S_{2}}_{R}(1k)=-\frac{1}{4\Lambda ^{5}}m_{l_{i}}g_{YB}(g_{B}+2g_{YB})[G_{6}(x_{\mu_{H}},x_{\tilde{R}_{j}},x_{\tilde{L}_{j}},x_{\tilde{L}_{i}},x_{B^{\prime}})(\sqrt{x_{B^{\prime}}}-\sqrt{x_{\mu_{H}}}\tan \beta)\nonumber\\
&&~~~~~~~~-2\sqrt{x_{B^{\prime}}}x_{\mu_{H}}G_{7}(x_{\mu_{H}},\tilde{x}_{R_{j}},x_{\tilde{L}_{j}},x_{\tilde{L}_{i}},x_{\tilde{B}^{\prime}})]\Delta_{jj}^{LR}\Delta_{ij}^{LL}E.
\end{eqnarray}

The one-loop contributions from Fig.\ref{Mia1}(1l)
\begin{eqnarray}
&&A^{S_{1}}_{L}(1l)=-\frac{1}{4\Lambda ^{5}}m_{l_{i}}g^{2}_{1}[G_{6}(x_{\mu_{H}},x_{\tilde{L}_{j}},x_{\tilde{R}_{j}},x_{\tilde{R}_{i}},x_{1})(\sqrt{x_{1}}-\sqrt{x_{\mu_{H}}}\tan \beta)\nonumber\\
&&~~~~~~~-2\sqrt{x_{1}}x_{\mu_{H}}G_{7}(x_{\mu_{H}},x_{\tilde{L}_{j}},x_{\tilde{R}_{j}},x_{\tilde{R}_{i}},x_{1})]\Delta_{jj}^{LR}\Delta_{ij}^{RR}E,\\
&&A^{S_{2}}_{L}(1l)=-\frac{1}{4\Lambda ^{5}}m_{l_{i}}g_{YB}(g_{B}+g_{YB})[G_{6}(x_{\mu_{H}},x_{\tilde{L}_{j}},x_{\tilde{R}_{j}},x_{\tilde{R}_{i}},x_{B^{\prime}})(\sqrt{x_{B^{\prime}}}-\sqrt{x_{\mu_{H}}}\tan \beta)\nonumber\\
&&~~~~~~~-2\sqrt{x_{B^{\prime}}}x_{\mu_{H}}G_{7}(x_{\mu_{H}},x_{\tilde{L}_{j}},x_{\tilde{R}_{j}},x_{\tilde{R}_{i}},x_{B^{\prime}})]\Delta_{jj}^{LR}\Delta_{ij}^{RR}E,\\
&&A^{S_{3}}_{L}(1l)=\frac{1}{4{\Lambda ^{5}}}m_{l_{i}}g^{2}_{2}[G_{6}(x_{\mu_{H}},x_{\tilde{L}_{j}},x_{\tilde{R}_{j}},x_{R_{i}},x_{2})(\sqrt{x_{2}}-\sqrt{x_{\mu_{H}}}\tan \beta)\nonumber\\
&&~~~~~~~-2\sqrt{x_{2}}x_{\mu_{H}}G_{7}(x_{\mu_{H}},x_{\tilde{L}_{j}},x_{\tilde{R}_{j}},x_{\tilde{R}_{i}},x_{2})]\Delta_{jj}^{LR}\Delta_{ij}^{RR}E.
\end{eqnarray}

The one-loop contributions from Fig.\ref{Mia1}(1m)
\begin{eqnarray}
&&A^{S_{1}}_{R}(1m)=-\frac{1}{2{\Lambda ^{5}}}m_{l_{j}}g^{2}_{1}[G_{6}(x_{\mu_{H}},x_{\tilde{L}_{j}},x_{\tilde{R}_{j}},x_{\tilde{R}_{i}},x_{1})(\sqrt{x_{1}}-\sqrt{x_{\mu_{H}}}\tan \beta)\nonumber\\
&&~~~~~~~-2\sqrt{x_{1}}x_{\mu_{H}}G_{7}(x_{\mu_{H}},x_{\tilde{L}_{j}},x_{\tilde{R}_{j}},x_{\tilde{R}_{i}},x_{1})]\Delta_{jj}^{LR}\Delta_{ij}^{RR}E,\\
&&A^{S_{2}}_{R}(1m)=-\frac{1}{4{\Lambda ^{5}}}m_{l_{j}}g_{YB}(g_{B}+2g_{YB})[G_{6}(x_{\mu_{H}},x_{\tilde{L}_{j}},x_{\tilde{R}_{j}},x_{\tilde{R}_{i}},x_{B^{'}})(\sqrt{x_{B^{\prime}}}-\sqrt{x_{\mu_{H}}}\tan \beta)\nonumber\\
&&~~~~~~~-2\sqrt{x_{B^{\prime}}}x_{\mu_{H}}G_{7}(x_{\mu_{H}},x_{\tilde{L}_{j}},x_{\tilde{R}_{j}},x_{\tilde{R}_{i}},x_{B^{\prime}})]\Delta_{jj}^{LR}\Delta_{ij}^{RR}E.
\end{eqnarray}

The one-loop contributions from Fig.\ref{Mia1}(1n)
\begin{eqnarray}
&&A^{S_{1}}_{L}(1n)=-\frac{1}{4\Lambda ^{5}}m_{l_{j}}g^{2}_{1}[G_{6}(x_{\mu_{H}},x_{\tilde{R}_{j}},x_{\tilde{L}_{j}},x_{L_{i}},x_{1})(\sqrt{x_{1}}-\sqrt{x_{\mu_{H}}}\tan \beta)\nonumber\\
&&~~~~~~~-2\sqrt{x_{1}}x_{\mu_{H}}G_{7}(x_{\mu_{H}},x_{\tilde{R}_{j}},x_{\tilde{L}_{j}},x_{\tilde{L}_{i}},x_{1})]\Delta_{jj}^{LR}\Delta_{ij}^{LL}E,\\
&&A^{S_{2}}_{L}(1n)=-\frac{1}{4\Lambda ^{5}}m_{l_{j}}g_{YB}(g_{B}+g_{YB})[G_{6}(x_{\mu_{H}},x_{\tilde{R}_{j}},x_{\tilde{L}_{j}},x_{\tilde{L}_{i}},x_{B^{\prime}})(\sqrt{x_{B^{\prime}}}-\sqrt{x_{\mu_{H}}}\tan \beta)\nonumber\\
&&~~~~~~~-2\sqrt{x_{B^{\prime}}}x_{\mu_{H}}G_{7}(x_{\mu_{H}},x_{\tilde{R}_{j}},x_{\tilde{L}_{j}},x_{\tilde{L}_{i}},x_{B^{\prime}})]\Delta_{jj}^{LR}\Delta_{ij}^{LL}E,\\
&&A^{S_{3}}_{L}(1n)=\frac{1}{4\Lambda ^{5}}m_{l_{j}}g^{2}_{2}[G_{6}(x_{\mu_{H}},x_{\tilde{R}_{j}},x_{\tilde{L}_{j}},x_{\tilde{L}_{i}},x_{2})(\sqrt{x_{2}}-\sqrt{x_{\mu_{H}}}\tan \beta)\nonumber\\
&&~~~~~~~-2\sqrt{x_{2}}x_{\mu_{H}}G_{7}(x_{\mu_{H}},x_{\tilde{R}_{j}},x_{\tilde{L}_{j}},x_{\tilde{L}_{i}},x_{2})]\Delta_{jj}^{LR}\Delta_{ij}^{LL}E.
\end{eqnarray}

The one-loop contributions from Fig.\ref{Mia2}(3a)
\begin{eqnarray}
&&A^{S_{1}}_{R}(3a)=\frac{1}{2\Lambda ^{2}}g^{2}_{1}\Delta_{ij}^{RR}C_{R}[G_{3}(x_{\tilde{R}_{j}},x_{1},x_{\tilde{R}_{i}})+G_{3}(x_{\tilde{R}_{i}},x_{1},x_{\tilde{R}_{j}})],\\
&&A^{S_{2}}_{R}(3a)=\frac{1}{8\Lambda ^{2}}(g_{B}+2g_{YB})^{2}\Delta_{ij}^{RR}C_{R}[G_{3}(x_{\tilde{R}_{j}},x_{B^{'}},x_{\tilde{R}_{i}})+G_{3}(x_{\tilde{R}_{i}},x_{B^{'}},x_{\tilde{R}_{j}})],\\
&&A^{S_{4}}_{L}(3a)=\frac{1}{4\Lambda ^{2}}Y_{e,ii}Y_{e,jj}C_{R}\Delta_{ij}^{RR}[G_{3}(x_{\tilde{R}_{j}},x_{\mu_{H}},x_{\tilde{R}_{i}})+G_{3}(x_{\tilde{R}_{i}},x_{\mu_{H}},x_{\tilde{R}_{j}})].
\end{eqnarray}

The one-loop contributions from Fig.\ref{Mia2}(3b)
\begin{eqnarray}
&&A^{S_{1}}_{L}(3b)=-\frac{1}{8\Lambda ^{2}}g^{2}_{1}\Delta_{ij}^{LL}C_{L}[G_{3}(x_{\tilde{L}_{j}},x_{1},x_{\tilde{L}_{i}})+G_{3}(x_{\tilde{L}_{i}},x_{1},x_{\tilde{L}_{j}})],\\
&&A^{S_{2}}_{L}(3b)=-\frac{1}{8\Lambda ^{2}}(g_{B}+g_{YB})^{2}\Delta_{ij}^{LL}C_{L}[G_{3}(x_{\tilde{L}_{j}},x_{B^{'}},x_{\tilde{L}_{i}})+G_{3}(x_{\tilde{L}_{i}},x_{B^{\prime}},x_{\tilde{L}_{j}})],\\
&&A^{S_{3}}_{L}(3b)=-\frac{1}{8\Lambda ^{2}}g^{2}_{2}C_{L}\Delta_{ij}^{LL}[G_{3}(x_{\tilde{L}_{j}},x_{2},x_{\tilde{L}_{i}})+G_{3}(x_{\tilde{L}_{i}},x_{2},x_{\tilde{L}_{j}})],\\
&&A^{S_{4}}_{R}(3b)=-\frac{1}{4\Lambda ^{2}}Y_{e,ii}Y_{e,jj}C_{L}\Delta_{ij}^{LL}[G_{3}(x_{\tilde{L}_{j}},x_{\mu_{H}},x_{\tilde{L}_{i}})+G_{3}(x_{\tilde{L}_{i}},x_{\mu_{H}},x_{\tilde{L}_{j}})].
\end{eqnarray}

The one-loop contributions from Fig.\ref{Mia2}(3c)
\begin{eqnarray}
&&A^{S_{1(2)}-S_{2(1)}}_{L}(3c)=\frac{1}{4\Lambda ^{2}}(\sqrt{x_{1}}+\sqrt{x_{B^{'}}})g_{1}(g_{B}+2g_{YB})\Delta_{ij}^{RR}\sqrt{x_{BB^{\prime}}}C_{R}\nonumber\\
&&~~~~~~~~~~~~~~\times[G_{4}(x_{\tilde{R}_{j}},x_{1},x_{\tilde{R}_{i}},x_{B^{\prime}})+G_{4}(x_{\tilde{R}_{i}},x_{1},x_{\tilde{R}_{j}},x_{B^{\prime}})].
\end{eqnarray}

The one-loop contributions from Fig.\ref{Mia2}(3d)
\begin{eqnarray}
&&A^{S_{1(2)}-S_{2(1)}}_{L}(3d)=-\frac{1}{8\Lambda ^{2}}(\sqrt{x_{1}}+\sqrt{x_{B^{\prime}}})g_{1}(g_{B}+g_{YB})\Delta_{ij}^{LL}\sqrt{x_{BB^{\prime}}}C_{L}\nonumber\\
&&~~~~~~~~~~~~~~\times[G_{4}(x_{\tilde{L}_{j}},x_{1},x_{\tilde{L}_{i}},x_{B^{\prime}})+G_{4}(x_{\tilde{L}_{i}},x_{1},x_{\tilde{L}_{j}},x_{B^{\prime}})].
\end{eqnarray}

The one-loop contributions from Fig.\ref{Mia2}(3e)
\begin{eqnarray}
&&A^{S_{1}}_{R}(3e)=\frac{1}{4\Lambda^{3}}(\sqrt{x_{\mu_{H}}}\tan \beta+\sqrt{x_{1}})m_{l_{j}}g^{2}_{1}\Delta^{LR}_{ij}\nonumber\\
&&~~~~~~~\times[C_{L}G_{4}(x_{\tilde{L}_{j}},x_{\mu_{H}},x_{\tilde{R}_{i}},x_{1})-C_{R}G_{4}(x_{\tilde{R}_{i}},x_{\mu_{H}},x_{\tilde{L}_{j}},x_{1})],\\
&&A^{S_{2}}_{R}(3e)=\frac{1}{8\Lambda^{3}}(\sqrt{x_{\mu_{H}}}\tan \beta+\sqrt{x_{B^{'}}})m_{l_{j}}g_{YB}(g_{B}+2g_{YB})\Delta^{LR}_{ij}\nonumber\\
&&~~~~~~~\times[C_{L}G_{4}(x_{\tilde{L}_{j}},x_{\mu_{H}},x_{\tilde{R}_{i}},x_{B^{\prime}})-C_{R}G_{4}(x_{\tilde{R}_{i}},x_{\mu_{H}},x_{\tilde{L}_{j}},x_{B^{\prime}})].
\end{eqnarray}

The one-loop contributions from Fig.\ref{Mia2}(3f)
\begin{eqnarray}
&&A^{S_{1}}_{L}(3f)=\frac{1}{8\Lambda ^{3}}(\sqrt{x_{\mu_{H}}}\tan \beta+\sqrt{x_{1}})m_{l_{j}}g^{2}_{1}\Delta^{LR}_{ij}\nonumber\\
&&~~~~~~~\times[C_{R}G_{4}(x_{\tilde{R}_{j}},x_{\mu_{H}},x_{\tilde{L}_{i}},x_{1})-C_{L}G_{4}(x_{\tilde{L}_{i}},x_{\mu_{H}},x_{\tilde{R}_{j}},x_{1})],\\
&&A^{S_{2}}_{L}(3f)=\frac{1}{8\Lambda ^{3}}(\sqrt{x_{\mu_{H}}}\tan \beta+\sqrt{x_{B^{\prime}}})m_{l_{j}}g_{YB}(g_{B}+2g_{YB})\Delta^{LR}_{ij}\nonumber\\
&&~~~~~~~\times[C_{R}G_{4}(x_{\tilde{R}_{j}},x_{\mu_{H}},x_{\tilde{L}_{i}},x_{B^{\prime}})-C_{L}G_{4}(x_{\tilde{L}_{i}},x_{\mu_{H}},x_{\tilde{R}_{j}},x_{B^{\prime}})],\\
&&A^{S_{3}}_{L}(3f)=-\frac{1}{8\Lambda ^{3}}(\sqrt{x_{\mu_{H}}}\tan\beta+\sqrt{x_{2}})m_{l_{j}}g^{2}_{2}\Delta_{ij}^{LR}\nonumber\\
&&~~~~~~~\times[C_{R}G_{4}(x_{\tilde{R}_{j}},x_{\mu_{H}},x_{\tilde{L}_{i}},x_{2})-C_{L}G_{4}(x_{\tilde{L}_{i}},x_{\mu_{H}},x_{\tilde{R}_{j}},x_{2})].
\end{eqnarray}

The one-loop contributions from Fig.\ref{Mia2}(3g)
\begin{eqnarray}
&&A^{S_{1}}_{R}(3g)=-\frac{1}{4\Lambda ^{3}}m_{l_{i}}g^{2}_{1}(\sqrt{x_{1}}+\sqrt{x_{\mu_{H}}}\tan \beta)\Delta_{ij}^{LR}\nonumber\\
&&~~~~~~~\times[C_{R}G_{4}(x_{\tilde{R}_{j}},x_{\mu_{H}},x_{\tilde{L}_{i}},x_{1})-C_{L}G_{4}(x_{\tilde{L}_{i}},x_{\mu_{H}},x_{\tilde{R}_{j}},x_{1})],\\
&&A^{S_{2}}_{R}(3g)=-\frac{1}{8\Lambda ^{3}}m_{l_{i}}g_{YB}(g_{B}+2g_{YB})(\sqrt{x_{B^{\prime}}}+\sqrt{x_{\mu_{H}}}\tan \beta)\Delta_{ij}^{LR}\nonumber\\
&&~~~~~~~\times[C_{R}G_{4}(x_{\tilde{R}_{j}},x_{\mu_{H}},x_{L_{i}},x_{B^{\prime}})-C_{L}G_{4}(x_{L_{i}},x_{\mu_{H}},x_{\tilde{R}_{j}},x_{B^{\prime}})].
\end{eqnarray}

The one-loop contributions from Fig.\ref{Mia2}(3h)
\begin{eqnarray}
&&A^{S_{1}}_{L}(3h)=-\frac{1}{8\Lambda ^{3}}m_{l_{i}}g^{2}_{1}(\sqrt{x_{1}}+\sqrt{x_{\mu_{H}}}\tan \beta)\Delta_{ij}^{LR}\nonumber\\
&&~~~~~~~\times[C_{L}G_{4}(x_{\tilde{L}_{j}},x_{\mu_{H}},x_{\tilde{R}_{i}},x_{1})-C_{R}G_{4}(x_{\tilde{R}_{i}},x_{\mu_{H}},x_{\tilde{L}_{j}},x_{1})],\\
&&A^{S_{2}}_{L}(3h)=-\frac{1}{8\Lambda ^{3}}m_{l_{i}}g_{YB}(g_{B}+g_{YB})(\sqrt{x_{B^{\prime}}}+\sqrt{x_{\mu_{H}}}\tan \beta)\Delta_{ij}^{LR}\nonumber\\
&&~~~~~~~\times[C_{L}G_{4}(x_{\tilde{L}_{j}},x_{\mu_{H}},x_{\tilde{R}_{i}},x_{B^{\prime}})-C_{R}G_{4}(x_{\tilde{R}_{i}},x_{\mu_{H}},x_{\tilde{L}_{j}},x_{B^{\prime}})],\\
&&A^{S_{3}}_{L}(3h)=\frac{1}{8\Lambda ^{3}}(\sqrt{x_{\mu_{H}}}\tan\beta+\sqrt{x_{2}})m_{l_{i}}g^{2}_{2}\Delta_{ij}^{LR}\nonumber\\
&&~~~~~~~\times[C_{L}G_{4}(x_{\tilde{L}_{j}},x_{\mu_{H}},x_{\tilde{R}_{i}},x_{2})-C_{R}G_{4}(x_{\tilde{R}_{i}},x_{\mu_{H}},x_{\tilde{L}_{j}},x_{2})].
\end{eqnarray}

The one-loop contributions from Fig.\ref{Mia2}(3i)
\begin{eqnarray}
&&A^{S_{1}}_{L}(3i)=-\frac{1}{8\Lambda ^{5}}(\sqrt{x_{\mu_{H}}}\tan \beta+\sqrt{x_{1}})m_{l_{j}}g^{2}_{1}\Delta_{jj}^{LR}\Delta_{ij}^{LL} \nonumber\\
&&~~~~~~~\times[C_{R}G_{6}(x_{\tilde{R}_{j}},x_{\mu_{H}},x_{\tilde{L}_{i}},x_{\tilde{L}_{j}},x_{1}) \nonumber\\
&&~~~~~~~-C_{L}\big(G_{6}(x_{\tilde{L}_{i}},x_{\mu_{H}},x_{\tilde{R}_{j}},x_{1})+G_{6}(x_{\tilde{L}_{j}}x_{\tilde{L}_{i}},x_{\mu_{H}},x_{\tilde{R}_{j}},x_{1})\big)],\\
&&A^{S_{2}}_{L}(3i)=-\frac{1}{8\Lambda ^{5}}(\sqrt{x_{\mu_{H}}}\tan \beta+\sqrt{x_{B^{\prime}}})m_{l_{j}}g_{YB}(g_{B}+g_{YB})\Delta_{jj}^{LR}\Delta_{ij}^{LL} \nonumber\\
&&~~~~~~~\times[C_{R}G_{6}(x_{\tilde{R}_{j}},x_{\mu_{H}},x_{\tilde{L}_{i}},x_{\tilde{L}_{j}}x_{B^{\prime}}) \nonumber\\
&&~~~~~~~-C_{L}\big(G_{6}(x_{\tilde{L}_{i}},x_{\mu_{H}},x_{\tilde{R}_{j}},x_{B^{\prime}})+G_{6}(x_{\tilde{L}_{j}}x_{\tilde{L}_{i}},x_{\mu_{H}},x_{\tilde{R}_{j}},x_{B^{\prime}})\big)],\\
&&A^{S_{3}}_{L}(3i)=\frac{1}{8\Lambda ^{5}}(\sqrt{x_{\mu_{H}}}\tan\beta+\sqrt{x_{2}})m_{l_{j}}g^{2}_{2}\Delta_{jj}^{LR}\Delta_{ij}^{LL} \nonumber\\
&&~~~~~~~\times[C_{R}G_{6}(x_{\tilde{R}_{j}},x_{\mu_{H}},x_{\tilde{L}_{i}},x_{\tilde{L}_{j}},x_{2}) \nonumber\\
&&~~~~~~~-C_{L}\big(G_{6}(x_{\tilde{L}_{i}},x_{\mu_{H}},x_{\tilde{R}_{j}},x_{2})+G_{6}(x_{\tilde{L}_{j}}x_{\tilde{L}_{i}},x_{\mu_{H}},x_{\tilde{R}_{j}},x_{2})\big)].
\end{eqnarray}

The one-loop contributions from Fig.\ref{Mia2}(3j)
\begin{eqnarray}
&&A^{S_{1}}_{R}(3j)=-\frac{1}{4\Lambda^{5}}(\sqrt{x_{\mu_{H}}}\tan \beta+\sqrt{x_{1}})m_{l_{j}}g^{2}_{1}\Delta^{LR}_{ii}\Delta^{LL}_{ij}\nonumber\\
&&~~~~~~~\times[C_{L}\big(G_{6}(x_{\tilde{L}_{j}},x_{\tilde{L}_{i}},x_{\mu_{H}},x_{\tilde{R}_{i}},x_{1})+G_{6}(x_{\tilde{L}_{i}},x_{\tilde{L}_{j}},x_{\mu_{H}},x_{\tilde{R}_{i}},x_{1})\big) \nonumber\\
&&~~~~~~~-C_{R}G_{6}(x_{\tilde{R}_{i}},x_{\mu_{H}},x_{\tilde{L}_{j}},x_{\tilde{L}_{i}},x_{1})],\\
&&A^{S_{2}}_{R}(3j)=-\frac{1}{8\Lambda^{5}}(\sqrt{x_{\mu_{H}}}\tan \beta+\sqrt{x_{B^{'}}})m_{l_{j}}g_{YB}(g_{B}+2g_{YB})\Delta^{LR}_{ii}\Delta^{LL}_{ij}\nonumber\\
&&~~~~~~~\times[C_{L}\big(G_{6}(x_{\tilde{L}_{j}},x_{\tilde{L}_{i}},x_{\mu_{H}},x_{\tilde{R}_{i}},x_{B^{\prime}})+G_{6}(x_{\tilde{L}_{i}},x_{\tilde{L}_{j}},x_{\mu_{H}},x_{\tilde{R}_{i}}x_{B^{\prime}})\big) \nonumber\\
&&~~~~~~~-C_{R}G_{6}(x_{\tilde{R}_{i}},x_{\mu_{H}},x_{\tilde{L}_{j}},x_{\tilde{L}_{i}}x_{B^{\prime}})].
\end{eqnarray}

The one-loop contributions from Fig.\ref{Mia2}(3k)
\begin{eqnarray}
&&A^{S_{1}}_{R}(3k)=\frac{1}{4\Lambda ^{5}}m_{l_{i}}g^{2}_{1}(\sqrt{x_{1}}+\sqrt{x_{\mu_{H}}}\tan \beta)\Delta_{jj}^{LR}\Delta_{ij}^{LL} \nonumber\\
&&~~~~~~~\times[C_{R}G_{6}(x_{\tilde{R}_{j}},x_{\mu_{H}},x_{\tilde{L}_{i}},x_{\tilde{L}_{j}},x_{1}) \nonumber\\
&&~~~~~~~-C_{L}\big(G_{6}(x_{\tilde{L}_{i}},x_{\mu_{H}},x_{\tilde{R}_{j}},x_{1})+G_{6}(x_{\tilde{L}_{j}}x_{\tilde{L}_{i}},x_{\mu_{H}},x_{\tilde{R}_{j}},x_{1})\big)],\\
&&A^{S_{2}}_{R}(3k)=\frac{1}{8\Lambda ^{5}}m_{l_{i}}g_{YB}(g_{B}+2g_{YB})(\sqrt{x_{B^{\prime}}}+\sqrt{x_{\mu_{H}}}\tan \beta)\Delta_{jj}^{LR}\Delta_{ij}^{LL}\nonumber\\
&&~~~~~~~\times[C_{R}G_{6}(x_{\tilde{R}_{j}},x_{\mu_{H}},x_{\tilde{L}_{j}},x_{\tilde{L}_{i}},x_{B^{\prime}}) \nonumber\\
&&~~~~~~~-C_{L}\big(G_{6}(x_{\tilde{L}_{i}},x_{\tilde{L}_{j}},x_{\mu_{H}},x_{\tilde{R}_{j}},x_{B^{\prime}})+G_{6}(x_{\tilde{L}_{j}},x_{\tilde{L}_{i}},x_{\mu_{H}},x_{\tilde{R}_{j}},x_{B^{\prime}}) \big)].
\end{eqnarray}

The one-loop contributions from Fig.\ref{Mia2}(3l)
\begin{eqnarray}
&&A^{S_{1}}_{L}(3l)=\frac{1}{8\Lambda ^{5}}m_{l_{i}}g^{2}_{1}(\sqrt{x_{1}}+\sqrt{x_{\mu_{H}}}\tan \beta)\Delta_{ii}^{LR}\Delta_{ij}^{LL} \nonumber\\
&&~~~~~~~\times[C_{L}\big(G_{6}(x_{\tilde{L}_{j}},x_{\tilde{L}_{i}},x_{\mu_{H}},x_{\tilde{R}_{i}},x_{1})+G_{6}(x_{\tilde{L}_{i}},x_{\tilde{L}_{j}},x_{\mu_{H}},x_{\tilde{R}_{i}},x_{1})\big) \nonumber\\
&&~~~~~~~-C_{R}G_{6}(x_{\tilde{R}_{i}},x_{\mu_{H}},x_{\tilde{L}_{j}},x_{\tilde{L}_{j}},x_{1})],\\
&&A^{S_{2}}_{L}(3l)=\frac{1}{8\Lambda ^{5}}m_{l_{i}}g_{YB}(g_{B}+g_{YB})(\sqrt{x_{B^{\prime}}}+\sqrt{x_{\mu_{H}}}\tan \beta)\Delta_{ii}^{LR}\Delta_{ij}^{LL} \nonumber\\
&&~~~~~~~\times[C_{L}\big(G_{6}(x_{\tilde{L}_{j}},x_{\tilde{L}_{i}},x_{\mu_{H}},x_{\tilde{R}_{i}},x_{B^{\prime}})+G_{6}(x_{\tilde{L}_{i}},x_{\tilde{L}_{j}},x_{\mu_{H}},x_{\tilde{R}_{i}},x_{B^{\prime}})\big) \nonumber\\
&&~~~~~~~-C_{R}G_{6}(x_{\tilde{R}_{i}},x_{\mu_{H}},x_{\tilde{L}_{j}},x_{\tilde{L}_{j}},x_{B^{\prime}})],\\
&&A^{S_{3}}_{L}(3l)=-\frac{1}{8\Lambda ^{5}}(\sqrt{x_{\mu_{H}}}\tan\beta+\sqrt{x_{2}})m_{l_{i}}g^{2}_{2}\Delta_{ii}^{LR}\Delta_{ij}^{LL} \nonumber\\
&&~~~~~~~\times[C_{L}\big(G_{6}(x_{\tilde{L}_{j}},x_{\tilde{L}_{i}},x_{\mu_{H}},x_{\tilde{R}_{i}},x_{2})+G_{6}(x_{\tilde{L}_{i}},x_{\tilde{L}_{j}},x_{\mu_{H}},x_{\tilde{R}_{i}},x_{2})\big) \nonumber\\
&&~~~~~~~-C_{R}G_{6}(x_{\tilde{R}_{i}},x_{\mu_{H}},x_{\tilde{L}_{j}},x_{\tilde{L}_{j}},x_{2})].
\end{eqnarray}

The one-loop contributions from Fig.\ref{Mia2}(3m)
\begin{eqnarray}
&&A^{S_{1}}_{R}(3m)=-\frac{1}{4\Lambda^{5}}(\sqrt{x_{\mu_{H}}}\tan \beta+\sqrt{x_{1}})m_{l_{j}}g^{2}_{1}\Delta^{LR}_{jj}\Delta^{RR}_{ij}\nonumber\\
&&~~~~~~~\times[C_{L}G_{6}(x_{\tilde{L}_{j}},x_{\tilde{R}_{j}},x_{\mu_{H}},x_{\tilde{R}_{i}},x_{1}) \nonumber\\
&&~~~~~~~-C_{R}\big(G_{6}(x_{\tilde{R}_{j}},x_{\mu_{H}},x_{\tilde{R}_{i}},x_{\tilde{L}_{j}},x_{1})+G_{6}(x_{\tilde{R}_{i}},x_{\mu_{H}},x_{\tilde{R}_{j}},x_{\tilde{L}_{j}},x_{1})\big)],\\
&&A^{S_{2}}_{R}(3m)=-\frac{1}{8\Lambda^{5}}(\sqrt{x_{\mu_{H}}}\tan \beta+\sqrt{x_{B^{'}}})m_{l_{j}}g_{YB}(g_{B}+2g_{YB})\Delta^{LR}_{jj}\Delta^{RR}_{ij}\nonumber\\
&&~~~~~~~\times[C_{L}G_{6}(x_{\tilde{L}_{j}},x_{\tilde{R}_{j}},x_{\mu_{H}},x_{\tilde{R}_{i}},x_{B^{\prime}}) \nonumber\\
&&~~~~~~~-C_{R}\big(G_{6}(x_{\tilde{R}_{j}},x_{\mu_{H}},x_{\tilde{R}_{i}},x_{\tilde{L}_{j}},x_{B^{\prime}})+G_{6}(x_{\tilde{R}_{i}},x_{\mu_{H}},x_{\tilde{R}_{j}},x_{\tilde{L}_{j}},x_{B^{\prime}})\big)].
\end{eqnarray}

The one-loop contributions from Fig.\ref{Mia2}(3n)
\begin{eqnarray}
&&A^{S_{1}}_{L}(3n)=-\frac{1}{8\Lambda ^{5}}(\sqrt{x_{\mu_{H}}}\tan \beta+\sqrt{x_{1}})m_{l_{j}}g^{2}_{1}\Delta_{ii}^{LR}\Delta_{ij}^{RR} \nonumber\\
&&~~~~~~~\times[C_{R}\big(G_{6}(x_{\tilde{R}_{i}},x_{\mu_{H}},x_{\tilde{L}_{i}},x_{\tilde{R}_{j}},x_{1})+G_{6}(x_{\tilde{R}_{j}},x_{\mu_{H}},x_{\tilde{L}_{i}},x_{\tilde{R}_{i}},x_{1})\big) \nonumber\\
&&~~~~~~~-C_{L}G_{6}(x_{\tilde{L}_{i}},x_{\tilde{R}_{i}},x_{\mu_{H}},x_{\tilde{R}_{j}},x_{1})],\\
&&A^{S_{2}}_{L}(3n)=-\frac{1}{8\Lambda ^{5}}(\sqrt{x_{\mu_{H}}}\tan \beta+\sqrt{x_{B^{\prime}}})m_{l_{j}}g_{YB}(g_{B}+g_{YB})\Delta_{ii}^{LR}\Delta_{ij}^{RR} \nonumber\\
&&~~~~~~~\times[C_{R}\big(G_{6}(x_{\tilde{R}_{i}},x_{\mu_{H}},x_{\tilde{L}_{i}},x_{\tilde{R}_{j}},x_{B^{\prime}})+G_{6}(x_{\tilde{R}_{j}},x_{\mu_{H}},x_{\tilde{L}_{i}},x_{\tilde{R}_{i}},x_{B^{\prime}})\big) \nonumber\\
&&~~~~~~~-C_{L}G_{6}(x_{\tilde{L}_{i}},x_{\tilde{R}_{i}},x_{\mu_{H}},x_{\tilde{R}_{j}},x_{B^{\prime}})],\\
&&A^{S_{3}}_{L}(3n)=\frac{1}{8\Lambda ^{5}}(\sqrt{x_{\mu_{H}}}\tan\beta+\sqrt{x_{2}})m_{l_{j}}g^{2}_{2}\Delta_{ii}^{LR}\Delta_{ij}^{RR} \nonumber\\
&&~~~~~~~\times[C_{R}\big(G_{6}(x_{\tilde{R}_{i}},x_{\mu_{H}},x_{\tilde{L}_{i}},x_{\tilde{R}_{j}},x_{2})+G_{6}(x_{\tilde{R}_{j}},x_{\mu_{H}},x_{\tilde{L}_{i}},x_{\tilde{R}_{i}},x_{2})\big) \nonumber\\
&&~~~~~~~-C_{L}G_{6}(x_{\tilde{L}_{i}},x_{\tilde{R}_{i}},x_{\mu_{H}},x_{\tilde{R}_{j}},x_{2})].
\end{eqnarray}

The one-loop contributions from Fig.\ref{Mia2}(3o)
\begin{eqnarray}
&&A^{S_{1}}_{R}(3o)=\frac{1}{4\Lambda ^{5}}m_{l_{i}}g^{2}_{1}(\sqrt{x_{1}}+\sqrt{x_{\mu_{H}}}\tan \beta)\Delta_{ii}^{LR}\Delta_{ij}^{RR} \nonumber\\
&&~~~~~~~\times[C_{R}\big(G_{6}(x_{\tilde{R}_{i}},x_{\mu_{H}},x_{\tilde{L}_{i}},x_{\tilde{R}_{j}},x_{1})+G_{6}(x_{\tilde{R}_{j}},x_{\mu_{H}},x_{\tilde{L}_{i}},x_{\tilde{R}_{i}},x_{1})\big) \nonumber\\
&&~~~~~~~-C_{L}G_{6}(x_{\tilde{L}_{i}},x_{\tilde{R}_{i}},x_{\mu_{H}},x_{\tilde{R}_{j}},x_{1})],\\
&&A^{S_{2}}_{R}(3o)=\frac{1}{8\Lambda ^{5}}m_{l_{i}}g_{YB}(g_{B}+2g_{YB})(\sqrt{x_{B^{\prime}}}+\sqrt{x_{\mu_{H}}}\tan \beta)\Delta_{ii}^{LR}\Delta_{ij}^{RR}\nonumber\\
&&~~~~~~~\times[C_{R}\big(G_{6}(x_{\tilde{R}_{i}},x_{\mu_{H}},x_{\tilde{L}_{i}},x_{\tilde{R}_{j}},x_{B^{\prime}})+G_{6}(x_{\tilde{R}_{j}},x_{\mu_{H}},x_{\tilde{L}_{i}},x_{\tilde{R}_{i}},x_{B^{\prime}})\big) \nonumber\\
&&~~~~~~~-C_{L}G_{6}(x_{\tilde{L}_{i}},x_{\tilde{R}_{i}},x_{\mu_{H}},x_{\tilde{R}_{j}},x_{B^{\prime}})].
\end{eqnarray}

The one-loop contributions from Fig.\ref{Mia2}(3p)
\begin{eqnarray}
&&A^{S_{1}}_{L}(3p)=\frac{1}{8\Lambda ^{5}}m_{l_{i}}g^{2}_{1}(\sqrt{x_{1}}+\sqrt{x_{\mu_{H}}}\tan \beta)\Delta_{jj}^{LR}\Delta_{ij}^{RR} \nonumber\\
&&~~~~~~~\times[C_{L}G_{6}(x_{\tilde{L}_{j}},x_{\tilde{R}_{j}},x_{\mu_{H}},x_{\tilde{R}_{i}},x_{1}) \nonumber\\
&&~~~~~~~-C_{R}\big(G_{6}(x_{\tilde{R}_{j}},x_{\mu_{H}},x_{\tilde{R}_{i}},x_{\tilde{L}_{j}},x_{1})+G_{6}(x_{\tilde{R}_{i}},x_{\mu_{H}},x_{\tilde{R}_{j}},x_{\tilde{L}_{j}},x_{1})\big)],\\
&&A^{S_{2}}_{L}(3p)=\frac{1}{8\Lambda ^{5}}m_{l_{i}}g_{YB}(g_{B}+g_{YB})(\sqrt{x_{B^{\prime}}}+\sqrt{x_{\mu_{H}}}\tan \beta)\Delta_{jj}^{LR}\Delta_{ij}^{RR} \nonumber\\
&&~~~~~~~\times[C_{L}G_{6}(x_{\tilde{L}_{j}},x_{\tilde{R}_{j}},x_{\mu_{H}},x_{\tilde{R}_{i}},x_{B^{\prime}}) \nonumber\\
&&~~~~~~~-C_{R}\big(G_{6}(x_{\tilde{R}_{j}},x_{\mu_{H}},x_{\tilde{R}_{i}},x_{\tilde{L}_{j}},x_{B^{\prime}})+G_{6}(x_{\tilde{R}_{i}},x_{\mu_{H}},x_{\tilde{R}_{j}},x_{\tilde{L}_{j}},x_{B^{\prime}})\big)],\\
&&A^{S_{3}}_{L}(3p)=-\frac{1}{8\Lambda ^{5}}(\sqrt{x_{\mu_{H}}}\tan\beta+\sqrt{x_{2}})m_{l_{i}}g^{2}_{2}\Delta_{jj}^{LR}\Delta_{ij}^{RR} \nonumber\\
&&~~~~~~~\times[C_{L}G_{6}(x_{\tilde{L}_{j}},x_{\tilde{R}_{j}},x_{\mu_{H}},x_{\tilde{R}_{i}},x_{2}) \nonumber\\
&&~~~~~~~-C_{R}\big(G_{6}(x_{\tilde{R}_{j}},x_{\mu_{H}},x_{\tilde{R}_{i}},x_{\tilde{L}_{j}},x_{2})+G_{6}(x_{\tilde{R}_{i}},x_{\mu_{H}},x_{\tilde{R}_{j}},x_{\tilde{L}_{j}},x_{2})\big)].
\end{eqnarray}

\bibliographystyle{plain}
\bibliography{a.bbl}

\begin{thebibliography}{99}

%\cite{Calibbi:2017uvl}
\bibitem{Calibbi:2017uvl}
L.~Calibbi and G.~Signorelli,
%``Charged Lepton Flavour Violation: An Experimental and Theoretical Introduction,''
Riv. Nuovo Cim. \textbf{41} (2018) no.2, 71-174
doi:10.1393/ncr/i2018-10144-0
[arXiv:1709.00294 [hep-ph]].
%249 citations counted in INSPIRE as of 21 Mar 2024

%\cite{Cabibbo:1963yz}
\bibitem{Cabibbo:1963yz}
N.~Cabibbo,
%``Unitary Symmetry and Leptonic Decays,''
Phys. Rev. Lett. \textbf{10} (1963), 531-533
doi:10.1103/PhysRevLett.10.531.
%7551 citations counted in INSPIRE as of 04 Apr 2024

%\cite{Kobayashi:1973fv}
\bibitem{Kobayashi:1973fv}
M.~Kobayashi and T.~Maskawa,
%``CP Violation in the Renormalizable Theory of Weak Interaction,''
Prog. Theor. Phys. \textbf{49} (1973), 652-657
doi:10.1143/PTP.49.652.
%11849 citations counted in INSPIRE as of 04 Apr 2024


%\cite{DoubleChooz:2011ymz}
\bibitem{DoubleChooz:2011ymz}
Y.~Abe \textit{et al.} [Double Chooz],
%``Indication of Reactor $\bar{\nu}_e$ Disappearance in the Double Chooz Experiment,''
Phys. Rev. Lett. \textbf{108} (2012), 131801
doi:10.1103/PhysRevLett.108.131801
[arXiv:1112.6353 [hep-ex]].
%1614 citations counted in INSPIRE as of 15 Nov 2023

%\cite{DayaBay:2012fng}
\bibitem{DayaBay:2012fng}
F.~P.~An \textit{et al.} [Daya Bay],
%``Observation of electron-antineutrino disappearance at Daya Bay,''
Phys. Rev. Lett. \textbf{108} (2012), 171803
doi:10.1103/PhysRevLett.108.171803
[arXiv:1203.1669 [hep-ex]].
%2833 citations counted in INSPIRE as of 15 Nov 2023



%\cite{Sun:2019wii}
\bibitem{Sun:2019wii}
K.~S.~Sun, J.~B.~Chen, X.~Y.~Yang and S.~K.~Cui,
%``LFV decays of Z boson in Minimal R-symmetric Supersymmetric Standard Model,''
Chin. Phys. C \textbf{43} (2019) no.4, 043101
doi:10.1088/1674-1137/43/4/043101
[arXiv:1901.03800 [hep-ph]].
%7 citations counted in INSPIRE as of 10 Jan 2024


%\cite{OLeary:2011vlq}
\bibitem{OLeary:2011vlq}
B.~O'Leary, W.~Porod and F.~Staub,
%``Mass spectrum of the minimal SUSY B-L model,''
JHEP \textbf{05} (2012), 042
doi:10.1007/JHEP05(2012)042
[arXiv:1112.4600 [hep-ph]].
%81 citations counted in INSPIRE as of 15 Nov 2023

%\cite{Abdallah:2014fra}
\bibitem{Abdallah:2014fra}
W.~Abdallah, S.~Khalil and S.~Moretti,
%``Double Higgs peak in the minimal SUSY B-L model,''
Phys. Rev. D \textbf{91} (2015) no.1, 014001
doi:10.1103/PhysRevD.91.014001
[arXiv:1409.7837 [hep-ph]].
%24 citations counted in INSPIRE as of 19 Nov 2023

%\cite{Barger:2008wn}
\bibitem{Barger:2008wn}
V.~Barger, P.~Fileviez Perez and S.~Spinner,
%``Minimal gauged U(1)(B-L) model with spontaneous R-parity violation,''
Phys. Rev. Lett. \textbf{102} (2009), 181802
doi:10.1103/PhysRevLett.102.181802
[arXiv:0812.3661 [hep-ph]].
%149 citations counted in INSPIRE as of 16 Nov 2023

%\cite{FileviezPerez:2008sx}
\bibitem{FileviezPerez:2008sx}
P.~Fileviez Perez and S.~Spinner,
%``Spontaneous R-Parity Breaking and Left-Right Symmetry,''
Phys. Lett. B \textbf{673} (2009), 251-254
doi:10.1016/j.physletb.2009.02.047
[arXiv:0811.3424 [hep-ph]].
%97 citations counted in INSPIRE as of 16 Nov 2023

%\cite{FileviezPerez:2010ek}
\bibitem{FileviezPerez:2010ek}
P.~Fileviez Perez and S.~Spinner,
%``The Fate of R-Parity,''
Phys. Rev. D \textbf{83} (2011), 035004
doi:10.1103/PhysRevD.83.035004
[arXiv:1005.4930 [hep-ph]].
%76 citations counted in INSPIRE as of 16 Nov 2023

%\cite{Nilles:1983ge}
\bibitem{Nilles:1983ge}
H.~P.~Nilles,
%``Supersymmetry, Supergravity and Particle Physics,''
Phys. Rept. \textbf{110} (1984), 1-162
doi:10.1016/0370-1573(84)90008-5.
%5811 citations counted in INSPIRE as of 19 Nov 2023

%\cite{Haber:1984rc}
\bibitem{Haber:1984rc}
H.~E.~Haber and G.~L.~Kane,
%``The Search for Supersymmetry: Probing Physics Beyond the Standard Model,''
Phys. Rept. \textbf{117} (1985), 75-263
doi:10.1016/0370-1573(85)90051-1.
%5423 citations counted in INSPIRE as of 19 Nov 2023

%\cite{Rosiek:1989rs}
\bibitem{Rosiek:1989rs}
J.~Rosiek,
%``Complete Set of Feynman Rules for the Minimal Supersymmetric Extension of the Standard Model,''
Phys. Rev. D \textbf{41} (1990), 3464
doi:10.1103/PhysRevD.41.3464.
%302 citations counted in INSPIRE as of 19 Nov 2023


%\cite{Aulakh:1999cd}
\bibitem{Aulakh:1999cd}
C.~S.~Aulakh, A.~Melfo, A.~Rasin and G.~Senjanovic,
%``Seesaw and supersymmetry or exact R-parity,''
Phys. Lett. B \textbf{459} (1999), 557-562
doi:10.1016/S0370-2693(99)00708-X
[arXiv:hep-ph/9902409 [hep-ph]].
%128 citations counted in INSPIRE as of 13 Jan 2024


%\cite{Yang:2010iq}
\bibitem{Yang:2010iq}
J.~M.~Yang,
%``Lepton flavor violating Z-boson decays at GigaZ as a probe of supersymmetry,''
Sci. China Phys. Mech. Astron. \textbf{53} (2010), 1949-1952
doi:10.1007/s11433-010-4146-3
[arXiv:1006.2594 [hep-ph]].
%16 citations counted in INSPIRE as of 10 Jan 2024

%\cite{Han:2011aq}
\bibitem{Han:2011aq}
X.~Han,
%``The Lepton Flavor Violating Decays $Z\to l_i l_j$ in the Simplest little Higgs Model,''
Mod. Phys. Lett. A \textbf{27} (2012), 1250158
doi:10.1142/S0217732312501581
[arXiv:1104.3534 [hep-ph]].
%4 citations counted in INSPIRE as of 10 Jan 2024

%\cite{Zhang:2014osa}
\bibitem{Zhang:2014osa}
H.~B.~Zhang, T.~F.~Feng, S.~M.~Zhao and F.~Sun,
%``Lepton flavor violation in the $\mu\nu$SSM with slepton flavor mixing,''
Int. J. Mod. Phys. A \textbf{29} (2014), 1450123
doi:10.1142/S0217751X14501231
[arXiv:1407.7365 [hep-ph]].
%15 citations counted in INSPIRE as of 15 Nov 2023





%\cite{Dong:2017ksc}
\bibitem{Dong:2017ksc}
X.~X.~Dong, S.~M.~Zhao, X.~J.~Zhan, Z.~J.~Yang, H.~B.~Zhang and T.~F.~Feng,
%``$Z\rightarrow l_i^{\pm}l_j^{\mp}$ processes in the BLMSSM,''
Chin. Phys. C \textbf{41} (2017) no.7, 073103
doi:10.1088/1674-1137/41/7/073103
[arXiv:1704.02202 [hep-ph]].
%1 citations counted in INSPIRE as of 15 Nov 2023

%\cite{Wang:2022tgf}
\bibitem{Wang:2022tgf}
Y.~T.~Wang, S.~M.~Zhao, T.~T.~Wang, X.~Wang, X.~X.~Long, J.~Ma and T.~F.~Feng,
%``Z boson decays Z\textrightarrow{}li\ensuremath{\pm}lj\ensuremath{\mp} and Higgs boson decays h\textrightarrow{}li\ensuremath{\pm}lj\ensuremath{\mp} with lepton flavor violation in a U(1) extension of the MSSM,''
Phys. Rev. D \textbf{106} (2022) no.5, 055044
doi:10.1103/PhysRevD.106.055044
[arXiv:2207.01770 [hep-ph]].
%3 citations counted in INSPIRE as of 15 Nov 2023

%\cite{Arganda:2015uca}
\bibitem{Arganda:2015uca}
E.~Arganda, M.~J.~Herrero, R.~Morales and A.~Szynkman,
%``Analysis of the h, H, A \textrightarrow{} \ensuremath{\tau}\ensuremath{\mu} decays induced from SUSY loops within the Mass Insertion Approximation,''
JHEP \textbf{03} (2016), 055
doi:10.1007/JHEP03(2016)055
[arXiv:1510.04685 [hep-ph]].
%57 citations counted in INSPIRE as of 16 Nov 2023

%\cite{Arganda:2016zvc}
\bibitem{Arganda:2016zvc}
E.~Arganda, M.~J.~Herrero, X.~Marcano, R.~Morales and A.~Szynkman,
%``Effective lepton flavor violating H\ensuremath{\ell}i\ensuremath{\ell}j vertex from right-handed neutrinos within the mass insertion approximation,''
Phys. Rev. D \textbf{95} (2017) no.9, 095029
doi:10.1103/PhysRevD.95.095029
[arXiv:1612.09290 [hep-ph]].
%39 citations counted in INSPIRE as of 16 Nov 2023

%\cite{Wang:2022iaf}
\bibitem{Wang:2022iaf}
T.~T.~Wang, S.~M.~Zhao, J.~F.~Zhang, X.~X.~Dong and T.~F.~Feng,
%``Lepton flavor violating decays $l_j\rightarrow {l_i\gamma }$ in the $U(1)_X$SSM model within the mass insertion approximation,''
Eur. Phys. J. C \textbf{82} (2022) no.7, 639
doi:10.1140/epjc/s10052-022-10613-5
[arXiv:2205.10485 [hep-ph]].
%0 citations counted in INSPIRE as of 16 Nov 2023

%\cite{Moroi:1995yh}
\bibitem{Moroi:1995yh}
T.~Moroi,
%``The Muon anomalous magnetic dipole moment in the minimal supersymmetric standard model,''
Phys. Rev. D \textbf{53} (1996), 6565-6575
doi:10.1103/PhysRevD.53.6565
[arXiv:hep-ph/9512396 [hep-ph]].
%649 citations counted in INSPIRE as of 20 Dec 2023



%\cite{Staub:2015kfa}
\bibitem{Staub:2015kfa}
F.~Staub,
%``Exploring new models in all detail with SARAH,''
Adv. High Energy Phys. \textbf{2015} (2015), 840780
doi:10.1155/2015/840780
[arXiv:1503.04200 [hep-ph]].
%281 citations counted in INSPIRE as of 20 Dec 2023




%\cite{Flores-Tlalpa:2001vbz}
\bibitem{Flores-Tlalpa:2001vbz}
A.~Flores-Tlalpa, J.~M.~Hernandez, G.~Tavares-Velasco and J.~J.~Toscano,
%``Effective Lagrangian description of the lepton flavor violating decays Z ---\ensuremath{>} l-+(i) l+-(j),''
Phys. Rev. D \textbf{65} (2002), 073010
doi:10.1103/PhysRevD.65.073010
[arXiv:hep-ph/0112065 [hep-ph]].
%35 citations counted in INSPIRE as of 16 Nov 2023


%\cite{ParticleDataGroup:2022pth}
\bibitem{ParticleDataGroup:2022pth}
R.~L.~Workman \textit{et al.} [Particle Data Group],
%``Review of Particle Physics,''
PTEP \textbf{2022} (2022), 083C01
doi:10.1093/ptep/ptac097.
%2334 citations counted in INSPIRE as of 22 Jan 2024


%\cite{Dedes:2015twa}
\bibitem{Dedes:2015twa}
A.~Dedes, M.~Paraskevas, J.~Rosiek, K.~Suxho and K.~Tamvakis,
%``Mass Insertions vs. Mass Eigenstates calculations in Flavour Physics,''
JHEP \textbf{06} (2015), 151
doi:10.1007/JHEP06(2015)151
[arXiv:1504.00960 [hep-ph]].
%29 citations counted in INSPIRE as of 21 Mar 2024


%\cite{Rosiek:2015jua}
\bibitem{Rosiek:2015jua}
J.~Rosiek,
%``MassToMI \textemdash{}A Mathematica package for an automatic Mass Insertion expansion,''
Comput. Phys. Commun. \textbf{201}, 144-158 (2016)
doi:10.1016/j.cpc.2015.12.011
[arXiv:1509.05030 [hep-ph]].
%16 citations counted in INSPIRE as of 04 Jan 2024

%\cite{Yang:2018guw}
\bibitem{Yang:2018guw}
J.~L.~Yang, T.~F.~Feng, Y.~L.~Yan, W.~Li, S.~M.~Zhao and H.~B.~Zhang,
%``Lepton-flavor violation and two loop electroweak corrections to $(g-2)_\mu$ in the B-L symmetric SSM,''
Phys. Rev. D \textbf{99} (2019) no.1, 015002
doi:10.1103/PhysRevD.99.015002
[arXiv:1812.03860 [hep-ph]].
%27 citations counted in INSPIRE as of 22 Mar 2024


%\cite{Dong:2024lvs}
\bibitem{Dong:2024lvs}
X.~X.~Dong, S.~M.~Zhao, J.~P.~Huo, T.~T.~Wang and T.~F.~Feng,
%``Charged lepton flavor violation in the B-L symmetric SSM,''
Phys. Rev. D \textbf{109}, no.5, 055019 (2024)
doi:10.1103/PhysRevD.109.055019
[arXiv:2402.19131 [hep-ph]].
%0 citations counted in INSPIRE as of 12 Apr 2024

%\cite{ATLAS:2019erb}
\bibitem{ATLAS:2019erb}
G.~Aad \textit{et al.} [ATLAS],
%``Search for high-mass dilepton resonances using 139 fb$^{-1}$ of $pp$ collision data collected at $\sqrt{s}=$13 TeV with the ATLAS detector,''
Phys. Lett. B \textbf{796} (2019), 68-87
doi:10.1016/j.physletb.2019.07.016
[arXiv:1903.06248 [hep-ex]].
%415 citations counted in INSPIRE as of 23 Jan 2024



%\cite{Cacciapaglia:2006pk}
\bibitem{Cacciapaglia:2006pk}
G.~Cacciapaglia, C.~Csaki, G.~Marandella and A.~Strumia,
%``The Minimal Set of Electroweak Precision Parameters,''
Phys. Rev. D \textbf{74} (2006), 033011
doi:10.1103/PhysRevD.74.033011
[arXiv:hep-ph/0604111 [hep-ph]].
%262 citations counted in INSPIRE as of 20 Jan 2024


%\cite{Carena:2004xs}
\bibitem{Carena:2004xs}
M.~Carena, A.~Daleo, B.~A.~Dobrescu and T.~M.~P.~Tait,
%``$Z^\prime$ gauge bosons at the Tevatron,''
Phys. Rev. D \textbf{70} (2004), 093009
doi:10.1103/PhysRevD.70.093009
[arXiv:hep-ph/0408098 [hep-ph]].
%621 citations counted in INSPIRE as of 20 Jan 2024

%\cite{Basso:2015xna}
\bibitem{Basso:2015xna}
L.~Basso,
%``The Higgs sector of the minimal SUSY $B-L$ model,''
Adv. High Energy Phys. \textbf{2015} (2015), 980687
doi:10.1155/2015/980687
[arXiv:1504.05328 [hep-ph]].
%23 citations counted in INSPIRE as of 20 Jan 2024

%\cite{Mahmoudi:2007gd}
\bibitem{Mahmoudi:2007gd}
F.~Mahmoudi,
%``New constraints on supersymmetric models from b ---\ensuremath{>} s gamma,''
JHEP \textbf{12} (2007), 026
doi:10.1088/1126-6708/2007/12/026
[arXiv:0710.3791 [hep-ph]].
%47 citations counted in INSPIRE as of 24 Jan 2024


%\cite{Olive:2008vv}
\bibitem{Olive:2008vv}
K.~A.~Olive and L.~Velasco-Sevilla,
%``Constraints on Supersymmetric Flavour Models from b ---\ensuremath{>} s gamma,''
JHEP \textbf{05} (2008), 052
doi:10.1088/1126-6708/2008/05/052
[arXiv:0801.0428 [hep-ph]].
%22 citations counted in INSPIRE as of 24 Jan 2024



%\cite{Muong-2:2021ojo}
\bibitem{Muong-2:2021ojo}
B.~Abi \textit{et al.} [Muon g-2],
%``Measurement of the Positive Muon Anomalous Magnetic Moment to 0.46 ppm,''
Phys. Rev. Lett. \textbf{126} (2021) no.14, 141801
doi:10.1103/PhysRevLett.126.141801
[arXiv:2104.03281 [hep-ex]].
%1653 citations counted in INSPIRE as of 22 Mar 2024

%\cite{MEG:2016leq}
\bibitem{MEG:2016leq}
A.~M.~Baldini \textit{et al.} [MEG],
%``Search for the lepton flavour violating decay $\mu ^+ \rightarrow \mathrm {e}^+ \gamma $ with the full dataset of the MEG experiment,''
Eur. Phys. J. C \textbf{76} (2016) no.8, 434
doi:10.1140/epjc/s10052-016-4271-x
[arXiv:1605.05081 [hep-ex]].
%1053 citations counted in INSPIRE as of 27 Mar 2024

%\cite{Belle:2021ysv}
\bibitem{Belle:2021ysv}
A.~Abdesselam \textit{et al.} [Belle],
%``Search for lepton-flavor-violating tau-lepton decays to $\ell\gamma$ at Belle,''
JHEP \textbf{10} (2021), 19
doi:10.1007/JHEP10(2021)019
[arXiv:2103.12994 [hep-ex]].
%63 citations counted in INSPIRE as of 27 Mar 2024


%\cite{ATLAS:2022uhq}
\bibitem{ATLAS:2022uhq}
G.~Aad \textit{et al.} [ATLAS],
%``Search for the charged-lepton-flavor-violating decay $Z\rightarrow e\mu$ in $pp$ collisions at $\sqrt{s}=13$ TeV with the ATLAS detector,''
Phys. Rev. D \textbf{108} (2023), 032015
doi:10.1103/PhysRevD.108.032015
[arXiv:2204.10783 [hep-ex]].
%13 citations counted in INSPIRE as of 22 Jan 2024

%\cite{ATLAS:2021bdj}
\bibitem{ATLAS:2021bdj}
G.~Aad \textit{et al.} [ATLAS],
%``Search for lepton-flavor-violation in $Z$-boson decays with $\tau$-leptons with the ATLAS detector,''
Phys. Rev. Lett. \textbf{127} (2022), 271801
doi:10.1103/PhysRevLett.127.271801
[arXiv:2105.12491 [hep-ex]].
%37 citations counted in INSPIRE as of 21 Mar 2024




\end{thebibliography}
\end{document}